\documentclass[onecolumn,amsmath,amssymb,floatfix,superscriptaddress]{revtex4-2}
\usepackage{dcolumn,graphicx,color,xcolor,caption,lineno,soul, ulem, array, comment}
\usepackage[colorlinks=true]{hyperref}

\newcommand{\ifn}{Istituto di Fotonica e Nanotecnologie - Consiglio Nazionale delle Ricerche (IFN-CNR), piazza Leonardo da Vinci 32, 20133 Milano, Italy}

\newcommand{\icfo}{ICFO - Institut de Ciencies Fotoniques, The Barcelona Institute of Science and Technology, 08860 Castelldefels (Barcelona), Spain}

\newcommand{\icrea}{ICREA – Institució Catalana de Recerca i Estudis Avançats, 08015 Barcelona, Spain}

\begin{document}
	
\title{Integrated thermo-optic phase shifters for laser-written photonic circuits operating at cryogenic temperatures}

\author{Francesco Ceccarelli}
\affiliation{\ifn}

\author{Jelena V. Rakonjac}
\affiliation{\icfo}

\author{Samuele Grandi}
\affiliation{\icfo}

\author{Hugues de Riedmatten}
\affiliation{\icfo}
\affiliation{\icrea}

\author{Roberto Osellame}
\affiliation{\ifn}

\author{Giacomo Corrielli}
\thanks{Corresponding author: giacomo.corrielli@cnr.it}
\affiliation{\ifn}
	
\begin{abstract}
Integrated photonics offers compact and stable manipulation of optical signals in miniaturized chips, with the possibility of changing dynamically their functionality by means of integrated phase shifters. Cryogenic operation of these devices is becoming essential for advancing photonic quantum technologies, accommodating components like quantum light sources, single photon detectors and quantum memories operating at liquid helium temperatures. In this work, we report on a programmable glass photonic integrated circuit (PIC) fabricated through femtosecond laser waveguide writing (FLW) and controlled by thermo-optic phase shifters both in a room-temperature and in a cryogenic setting. By taking advantage of a femtosecond laser microstructuring process, we achieved reliable PIC operation with minimal power consumption and confined temperature gradients in both conditions. This advancement marks the first cryogenically-compatible programmable FLW PIC, paving the way for fully integrated quantum architectures realized on a laser-written photonic chip.
\end{abstract}

\maketitle
 
\section{Introduction}
Integrated photonics is an enabling technology that plays a central role in a growing number of applications, from signal routing and optical interconnects in high-speed data centers \cite{havemann2001high,beausoleil2011large}, to optical computing \cite{shastri2021photonics} and quantum information processing \cite{wang2020integrated}. Waveguide-based photonic integrated circuits (PICs) allow for the coherent manipulation of optical signals in an interferometrically stable fashion and with a miniaturized footprint, even when a large number of components are cascaded. Furthermore, PICs can incorporate integrated light modulators, e.g. by exploiting the thermo-optic or the electro-optic response of the waveguides, thus enabling dynamical programming of their function by means of external electrical signals. All these features make PICs capable of carrying out complex photonic operations that are otherwise impossible.

In recent years, the need for PICs operating at cryogenic temperatures has emerged, largely driven by the rapid development of photonic quantum technologies. Indeed, several essential quantum building blocks, including single-emitter-based quantum light sources \cite{davanco2017heterogeneous}, solid state quantum memories \cite{zhou2023photonic}, spin-photon interfaces \cite{xiong2023high} and superconducting nanowire single-photon detectors (SNSPDs) \cite{esmaeil2021superconducting}, work from few K down to the mK regime, and their integration in quantum photonic processors requires PICs operating reliably also at this temperature range. In addition, cryogenic PICs could also simplify qubit addressing in trapped-ions quantum computers \cite{mehta2016integrated, timpu2022laser}, or could reduce the heat load associated to microwave field delivery in superconducting quantum processors \cite{lecocq2021control}.

The first demonstrations of cryogenically compatible PICs manufactured by silicon-on-insulator (SOI), silicon nitride (SiN) and lithium niobate (LN) planar technologies have been recently reported, integrating electro-optic \cite{pintus2019characterization, eltes2020integrated, chakraborty2020cryogenic, lee2020high, youssefi2021cryogenic, lomonte2021single, gehl2017operation}, magneto-optic \cite{pintus2022integrated} or elasto-optic modulators \cite{dong2021cryogenically}. Another successful strategy for obtaining programmable PICs working at low temperature consists in the integration of micro-electromechanical systems (MEMS) within SiN PICs for modulating their functioning via a mechanical deformation of specific waveguide segments \cite{beutel2022cryo, gyger2021reconfigurable}. These devices offer striking speed performances, enabling light amplitude/phase modulation with a bandwidth ranging from a few MHz for MEMS-based devices up to tens of GHz for electro-optic components, and almost null static power dissipation. However, their fabrication relies on the heterogeneous integration of different materials, e.g. barium titanate \cite{lee2020high}, cerium-substituted yttrium iron garnet \cite{pintus2022integrated} or aluminum nitride \cite{dong2021cryogenically}, and typically requires highly sophisticated manufacturing processes. In addition, these kinds of integrated modulators suffer from increased optical losses, and their integration with coherent emitters and quantum memory materials is not straightforward. Due to these reasons, the scalability of these approaches and their applicability to photonic quantum technologies is still under study.

In the case of room-temperature quantum photonic experiments, a common solution for manufacturing reconfigurable PICs consists of using thermo-optic phase shifters (TOPSs), which exploit integrated resistive heaters for locally altering the waveguide effective index via Joule effect. Despite being slower, with bandwidths that reach a few hundreds of kHz in best cases \cite{harris2014efficient, tong2023efficient}, TOPSs can be easily integrated at a large scale of up to thousands within a single device \cite{suzuki2019low}, with no significant increase of the chip optical losses. However, realising quantum PICs based on TOPSs and operating reliably at cryogenic temperatures is challenging for two main reasons. First, the thermo-optic coefficient of relevant materials for PIC manufacturing drops by several orders of magnitude at cryogenic temperatures \cite{chakraborty2020cryogenic, komma2012thermo, elshaari2016thermo}. Second, heat dissipation and thermal load management inside a cryostat are non-trivial tasks, especially when trying to avoid temperature gradients at the PIC regions where the temperature-sensitive devices, e.g. SNSPDs or quantum memories, are located. Due to these factors, cryogenically compatible TOPS demonstrations have been limited thus far \cite{elshaari2016thermo, han2023cryogenic}.

In this work we fabricated a programmable glass PIC by the femtosecond laser waveguide writing (FLW) technique, actuated by means of different TOPSs with variable geometry, and we characterized it both at room temperature and in a cryogenic environment. We show that, in agreement with previous studies from our group \cite{ceccarelli2020low}, a suitable substrate microstructuring quenches thermal diffusion outside the TOPS region and produces a reliable PIC operation with low power dissipation ($< 3$ mW for full $2\pi$ phase modulation) in both temperature regimes.
FLW glass circuits are a valuable tool in the development of photonic quantum technologies \cite{corrielli2021femtosecond}, as they show low losses, excellent connectivity with optical fibers, and can include a large number of TOPSs to produce arbitrary unitary linear photonic operations \cite{dyakonov2018reconfigurable, pentangelo2024high}. In addition, FLW PICs are compatible with quantum-dot single photon sources \cite{anton2019interfacing}, and can integrate both quantum memories \cite{rakonjac2022storage, zhang2023telecom}, diamond SiV centers \cite{koch2022super} and SNSPDs \cite{han2023cryogenic}. Our result represents the first example of a programmable FLW PIC operating at cryogenic temperatures, thus paving the way to the realization of a complete quantum architecture, fully integrated on a FLW chip.

\section{Methods}
\subsection{Device fabrication}

The PIC under study contains three independent Mach-Zehnder interferometers (MZIs), each one encompassing a TOPS on one arm for controlling the internal phase, and thus behaving as an integrated light intensity modulator. For each MZI we varied the TOPS length, in order to study the dependence of its efficiency from this geometrical parameter. The PIC was entirely fabricated by FLW in glass following a three-step process, consisting in: i) direct laser writing of the optical waveguides composing the integrated MZIs; ii) 3D laser ablation of the glass substrate around the waveguides for improving heat confinement during phase shifting operation; iii) deposition of a metal layer on top of the glass substrate and definition of the resistors by laser ablation of the metal layer. In the following we discuss this process in detail.\\

\begin{figure}[h]
	\centering
	\includegraphics[width=1\linewidth]{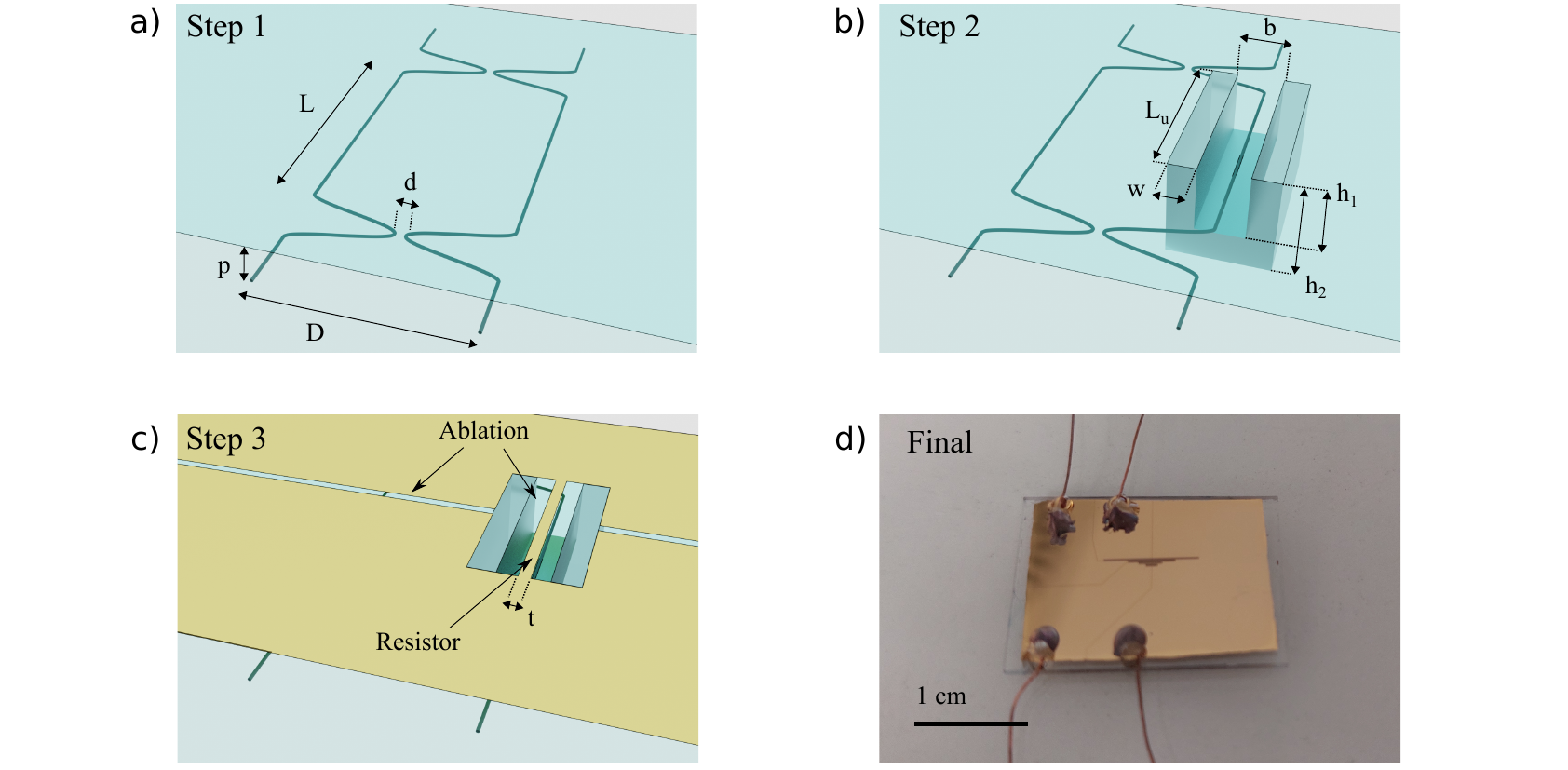}
	\caption{(a) Schematic of the fabricated MZIs. The DCs have been fabricated by concatenating different arches of circumference of radius $R$~=~35~mm, which bring the two waveguides of the DC at a minimum interaction distance $d$~=~4.65~$\mu$m. Waveguide separation $D$ before and after the DCs is 100 $\mu$m. Waveguides distance $p$ from the top surface is 50~$\mu$m. Three different MZIs have been fabricated with different lengths $L$ of the interferometric arms (1~mm, 5~mm and 10~mm). (b) Schematic of the geometry of the 3D undercuts. For all MZIs: $b$~=~30~$\mu$m, $W$~=~70~$\mu$m, $h_1$~=~75~$\mu$m and $h_2$~=~120~$\mu$m. The length of the MZI arms differ for the values of $L_u$, which are 1~mm, 4~mm and 8~mm. (c) Schematic of the metal film patterning. Laser ablation is used to narrow the lateral width of the resistive heaters to the value of $t$~=~15~$\mu$m and to isolate large pads for electrical connections. (d) Photograph of the device at the end of the fabrication process.}
	\label{Fig1}
\end{figure}

\textbf{Step 1.} High quality optical waveguides have been fabricated by focusing a femtosecond-pulsed laser beam, produced by a homemade Yb:KYW cavity-dumped laser oscillator (wavelength of 1030~nm, pulse duration of 300~fs, repetition rate of 1~MHz) inside a 1~mm thick substrate of a commercial boro-aluminosilicate glass (Eagle~XG, from Corning). The focusing optic was a 50x microscope objective with numerical aperture of 0.65. Each waveguide was fabricated by overlapping 12 laser scans (pulse energy of 230~nJ) performed at the speed of 40~mm/s. After laser irradiation, the sample was annealed at a maximum temperature of 750~°C for a total annealing duration of 24~h (full annealing process described in \cite{nayak2021first}). The waveguides fabricated in this way show single mode behavior at 606~nm wavelength ($1/e^2$ mode diameter of $\approx$~2.5~$\mu$m), propagation losses $<$~0.2~dB/cm and negligible bending losses for bending radii $>$~30~mm, with no appreciable differences for TE and TM polarized light.
Using this fabrication recipe, we inscribed in the sample a set of three MZIs, each one obtained by cascading two 50/50 directional couplers (DCs), according to the geometry presented in figure \ref{Fig1}(a). Each MZI was inscribed at a depth $p =$  50 $\mu$m, and differ from each other by the length $L$ of the straight waveguide segments between the DCs which form the MZI arms. The values of $L$ implemented are 1~mm, 5~mm and 10~mm. The lateral facets of the sample were polished to optical quality in order to expose the waveguide terminations and enable light coupling.\\

\textbf{Step 2.} After the fabrication of the photonic circuit, we proceeded with the glass microstructuring around the waveguides, with the goal of creating a 3D heat confining structure for enhancing the efficiency of the thermal phase shifters. To do so, we employed the water-assisted laser ablation technique, which consists of irradiating the sample with focused laser pulses for obtaining direct material removal. This process is performed with the sample immersed in deionized water, which helps to remove the glass debris produced during the machining. With this technique we fabricated deep undercuts around one arm of each MZI, according to the geometry shown in figure \ref{Fig1}(b), which leaves the waveguides confined in a suspended glass bridge with a rectangular cross section. For the ablation process we used a train of laser pulses (1030 nm wavelength, 1~ps duration, 20~kHz repetition rate, 1.1~$\mu$J/pulse) from a commercial source (Pharos, from Light Conversion), focused with a water-immersion microscope objective with 0.5 numerical aperture. We fabricated the undercuts with three different longitudinal lengths $L_u$ of 1~mm, 4~mm and 8~mm. For the structures longer than 1~mm, we also added small reinforcement pillars (one every 1~mm, 30 $\mu$m $\times$ 50 $\mu$m rectangular cross section) that connect the suspended glass bridge to the bottom of the undercut.\\

\textbf{Step 3.} The TOPSs were fabricated according to the process reported in \cite{ceccarelli2020low}. A Cr/Au (5/100 nm) film was deposited on the glass substrate with a thermal evaporation process (Moorfield MINILAB-080) and then an annealing step (400 \textdegree C for 1 h) was employed to prevent long term drifts of the resistance value (and thus of the induced phase) during the operation at high temperature. The sheet resistance of the Cr/Au metal film at the end of this process is $R_s~\approx~$~2.0~$\Omega/\square$. After this, we used again FLW for patterning the metal film to define the resistive heaters on top of the suspended waveguides, and to isolate the electrical pads for their connectorization and control (see figure \ref{Fig1}(c)). To do so we employed the same laser used in step 2, with different settings: 170 fs pulse duration, 1 MHz repetition rate, 200 nJ pulse energy, 0.25 numerical aperture focusing optics. The heaters fabricated in this way have the same length $L_u$ of the corresponding undercut, and a lateral width $t$ of 15 $\mu$m.
Finally, we glued four copper wires (three heaters contacts and a common ground) to the metal film at the electrical pads using both an electrically conductive epoxy and a UV curing transparent glue. A photograph of the final device can be seen in figure \ref{Fig1}(d).

\subsection{Characterization setup}
\begin{figure}[t]
	\centering
	\includegraphics[width=1\linewidth]{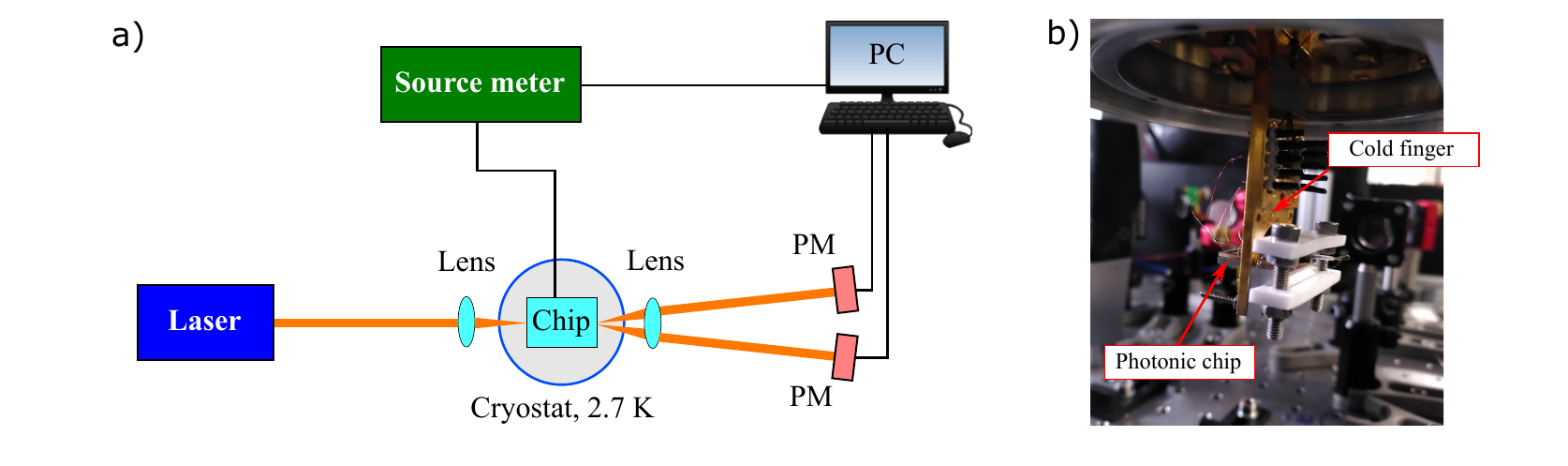}
	\caption{(a) Sketch of the experimental setup used for the PIC characterization. The lenses used for in- and out-coupling of the light in the chip have a focal length of 10 cm. (b) Photograph of the PIC mounted on the cold finger, before the closure of the cryostat.}
	\label{Fig2}
\end{figure}

The setup used for the characterization of the PIC is shown in figure \ref{Fig2}(a). The PIC was mounted inside a closed-cycle pulse-tube cryostat (Optistat AC-V14, Oxford Instruments; cooling power 0.5 W @ 4 K). The operating pressure inside the sample chamber of the cryostat is $\mathrm{p_C}=9.3\cdot 10^{-6}$ mbar, and the cold finger reaches a minimum temperature $T_C$ = 2.7 K. The PIC itself was glued to an aluminium mount with thermally conductive varnish (GE 7031), and the mount was bolted to the cold finger with Apiezon N grease between the two parts for increased thermal contact. A photograph of the PIC mounted in the cryostat (taken before the closure of the cryogenic chamber) is shown in figure \ref{Fig2}(b). Continuous-wave laser light at 606 nm, provided by a commercial laser source (Toptica TA-SHG), is free-space coupled in and out the photonic chip through two lateral viewports mounted on the side walls of the cryogenic chamber, by means of two plano-convex lenses. Two powermeters (PMs) allowed to measure the optical power at the output ports of each MZI simultaneously. The actuation of the TOPSs was performed by means of a programmable sourcemeter (Keysight B2902A), which allows to set and measure electrical currents and voltages with source resolution down to 1 pA/1 $\mu$V and measurement resolution down to 100 fA/100 nV. Both the sourcemeter and the PMs were connected to a PC, for performing automated measurements.\\

\section{Results and discussion}\label{Sec_Results}
\subsection{Devices characterization}
\begin{figure}[t]
	\centering
	\includegraphics[width=1\linewidth]{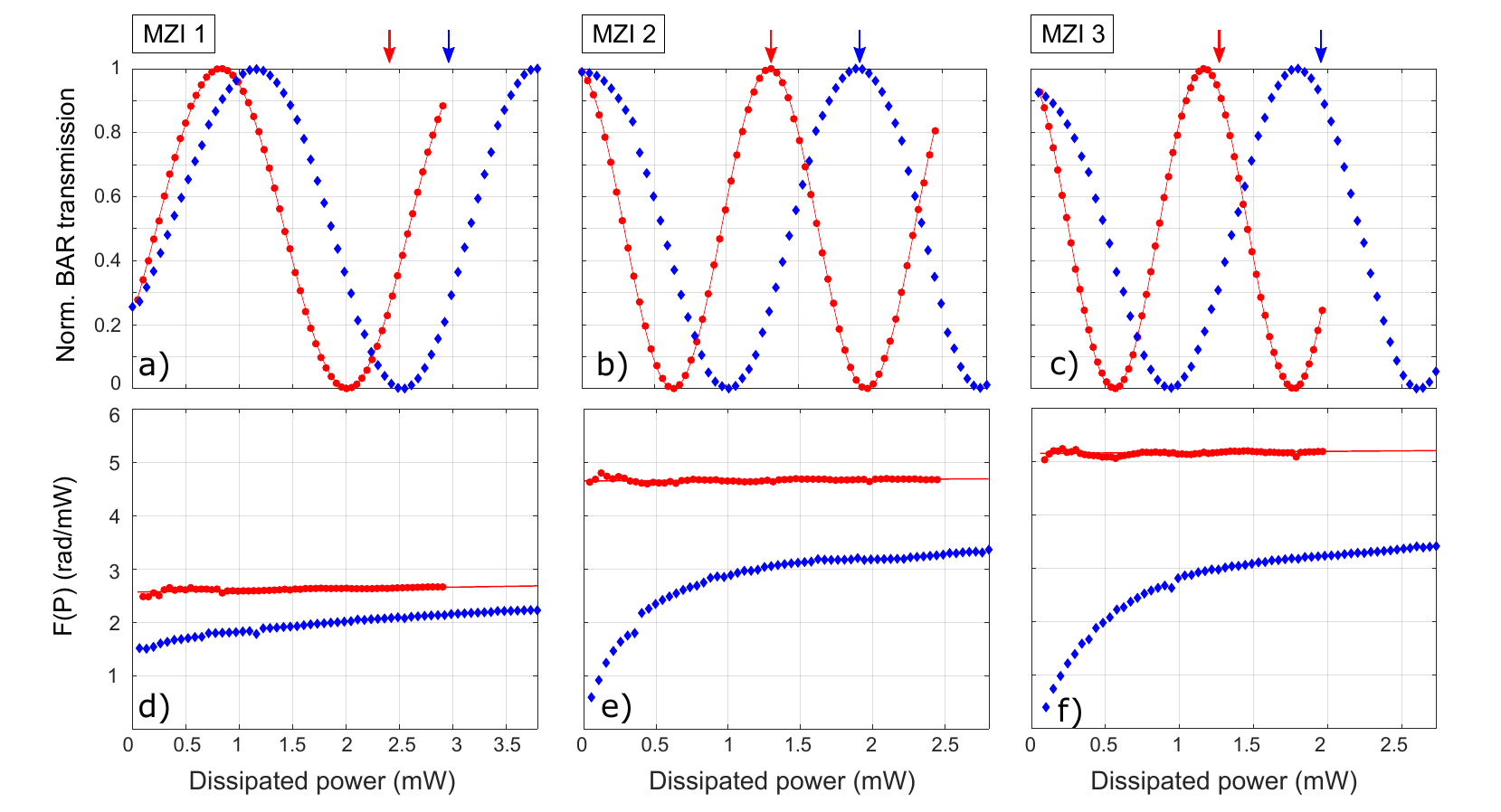}
	\caption{(a-c) Experimental characterization of $\mathcal{T_{BAR}}$ at room (red dots) and cryogenic (blue diamonds) temperature. The three plots (a), (b) and (c) correspond to MZI 1, 2 and 3 (see table \ref{Tab1}),  respectively. The solid red line is best fit function of room temperature data, according to equations \ref{eqBar} and \ref{eqFi}. Red and blue arrows indicate the value of dissipated power needed to attain a $2\pi$ phase shift in the two cases. (d-f) Computed trends of the TOPS efficiencies at room (red dots) and cryogenic (blue diamonds) temperatures as a function of dissipated power, according to equation \ref{eqInverted}. The solid red line is a plot of $\alpha+\beta P$, according to best fit parameters listed in table \ref{Tab1}.}
	\label{Fig3}
\end{figure}
A first characterization of the PIC was performed at room temperature (297.4 K), with the cryostat closed and the sample kept at the operating pressure $\mathrm{p_C}$. We coupled the laser light at one input of each MZI and we measured the output powers $\mathcal{P}_\mathrm{BAR}$ and $\mathcal{P}_\mathrm{CROSS}$ at both bar and cross output ports simultaneously, while driving electrical current to the TOPSs. From these measurements we computed the normalized bar transmission $\mathcal{T}_\mathrm{BAR}~=~\mathcal{P}_\mathrm{BAR}/(\mathcal{P}_\mathrm{BAR}~+~\mathcal{P}_\mathrm{CROSS})$ as a function of the dissipated electrical power $P$ for each MZI. These results are shown in figures \ref{Fig3}(a-c), red dots. For the data fitting (red solid line) we used the standard transmission model of a balanced MZI with unit visibility:

\begin{equation}\label{eqBar}
    \mathcal{T}_\mathrm{BAR}(P) = \frac{1}{2}-\frac{1}{2}\mathrm{cos}\left(\phi_0+\phi(P)\right).
\end{equation}

Here, $\phi_0$ represents the static optical phase difference between the MZI arms due to fabrication tolerances, while $\phi(P)$ is the optical phase difference induced by heat dissipation as function of $P$, that, in the case of room temperature operation, is well described by the following relation:

\begin{equation}\label{eqFi}
    \phi(P)=\alpha P + \beta P^2.
\end{equation}

In this expression, $\alpha$ represents the linear efficiency of the TOPS, while $\beta$ is a non-linear correction coefficient that accounts for the temperature dependence of the relevant glass thermal parameters, i.e. the thermo-optic coefficient $c_{to}$ and the thermal conductivity $k_\textup{glass}$. Even though at room temperature this dependence is very weak and might be ignored in the first place, we decided to include it in our data analysis in order to get a more robust extraction of the static phase $\phi_0$, which will be used for the subsequent analysis of the cryogenic behaviour. The values of the fitted parameters for all MZIs are listed in table \ref{Tab1}. From these measurements we could also retrieve the values of dissipated power $P^A_{2\pi}$ for inducing a $2\pi$ phase shift in the case of room temperature operation (also listed in table \ref{Tab1}, and indicated in figures \ref{Fig3}(a-c) with a red arrow), which range from 1.27 mW to 2.40 mW. Such low values are the result of the strong thermal insulation of the glass bridge containing the actuated waveguide. They are significantly lower, by more than an order of magnitude, compared to the usual results achieved under standard pressure conditions and are in good agreement with previous characterizations of similar devices in vacuum environment \cite{ceccarelli2020low}. It is also worth noting that, differently from the case of operation in standard pressure conditions \cite{ceccarelli2020low}, the length of the TOPS plays an important role in determining the efficiency (and thus the power dissipation) of the device, where longer TOPSs are more efficient than the shorter one. 

We then proceeded by characterizing the MZIs with the cryostat turned on, with the cold finger sensor reaching the temperature $T_C$. An additional thermal probe glued on top of the PIC reported a steady temperature of $T_P$~=~16.5~K, thus allowing us to bound the actual temperature of the PIC between $T_C$ and $T_P$. We attribute this discrepancy to imperfect thermal contact between the PIC and the cold finger and/or between the PIC and the thermal probe. In this condition, we repeated the measurement of $\mathcal{T_{BAR}}(P)$ for all MZIs, and the results are plotted in figure \ref{Fig3}(a-c), blue diamonds. Also in this regime we were able to record complete oscillation fringes in the MZI bar transmission, with an observed increase of power dissipation for $2\pi$ phase shift of $\approx$ 20\% for the TOPS with length of 1~mm, and $\approx$~50\% for the TOPS with lengths of 4~mm and 8~mm. The actual measured values $P^B_{2\pi}$, indicated with blue arrows in figure \ref{Fig3}(a-c), are reported in table \ref{Tab1}. Importantly, no significant temperature variation at the temperature sensors had been observed during the TOPS actuation in both temperature regimes.

Unlike the room-temperature case, the experimental data acquired at cryogenic temperatures are no longer correctly modelled by equation \ref{eqFi}. In particular, the functional shape of $\mathcal{T_{BAR}}(P)$ is still described by equation \ref{eqBar}, but shows a strong and non-linear frequency chirp that reduces the TOPS efficiency for lower values of dissipated power. In order to better quantify this effect and to obtain a direct comparison with the room temperature behavior, we assumed a more generic model for the dependence of $\phi(P)$:
\begin{equation}\label{eqFiGeneric}
    \phi(P)=F(P)\cdot P,
\end{equation}
where $F(P)$ represents the TOPS linear efficiency as a function of the dissipated power. We computed this quantity by inverting numerically the experimental values of $\mathcal{T_{BAR}(P)}$, according to the formula:
\begin{equation}\label{eqInverted}
    F(P)= \frac{\mathrm{cos}^{-1}\left(1-2\mathcal{T_{BAR}}(P)\right)-\phi_0}{P}.
\end{equation}
In doing this operation we used the values of $\phi_0$ retrieved from the room temperature data fit (see table \ref{Tab1}). The results are shown in figure \ref{Fig3}(d-f) for all MZIs, computed from both room-temperature (red dots) and cryogenic (blue diamonds) characterization. From these plots one can appreciate that, in the case of room temperature operation, the TOPS efficiency is essentially constant and equal to $\alpha$. At cryogenic temperatures the efficiency becomes smaller, and drops significantly at low power dissipation values, for which the operating temperature of the TOPSs is closer to $T_C$. This effect is more pronounced for the longer TOPSs. We attribute this behavior to a significant decrease of the thermo-optic coefficient of Eagle XG glass at low temperature, an effect similar to the one already reported in the literature for fused silica, silicon and silicon nitride \cite{komma2012thermo, elshaari2016thermo}.

\begin{table}[t]
    \centering
    \begin{tabular}{|m{1.5cm}|m{1.5cm}|m{1.5cm}|m{1.5cm}|m{2cm}|m{2cm}|m{1.5cm}|m{1.5cm}|m{1.5cm}|}
    \hline
        Device & $L_u$ (mm) & $P_{2\pi}^A$ (mW) & $P_{2\pi}^B$ (mW) & $\alpha$ (rad/mW) & $\beta$ (rad/mW$^2$) & $\phi_0$ (rad) & $\Delta T^{2\pi}_M$ (K) & $c^*_{to}$ (K$^{-1}$) \\
        \hline
        MZI 1 & 1 & 2.40 & 2.95 & 2.57 & 0.033 & 0.99 & 71.7 & 7.0E-6 \\
        \hline
        MZI 2 & 4 & 1.35 & 1.98 & 4.65 & 0.014 & 3.35 & 19.9 & 7.3E-6 \\
        \hline
        MZI 3 & 8 & 1.27 & 1.94 & 5.15 & 0.020 & 3.40 & 9.4 & 7.9E-6 \\
        \hline
    \end{tabular}
    \caption{Summary of the characterization of the different MZIs. $L_u$ indicates the length of the TOPS. $P_{2\pi}$ indicates the value of electrical power required to produce a $2\pi$ phase shift at room (A) and cryogenic (B) temperatures. $\alpha$, $\beta$ and $\phi_0$ are the fit parameters according to the model described by equations \ref{eqBar} and \ref{eqFi}, applied to case (A). $\Delta T_M^{2\pi}$ indicates the value of mean temperature increase inside the TOPS for $P=P_{2\pi}^A$, retrieved from numerical simulations (see figure \ref{Fig4}(b)). $c^*_{to}$ is the estimated value of the thermo-optic coefficient of Eagle XG at room temperature.}
    \label{Tab1}
\end{table}

\subsection{Numerical simulations and discussion}
\begin{figure}[t]
    \centering
    \includegraphics[width=1\linewidth]{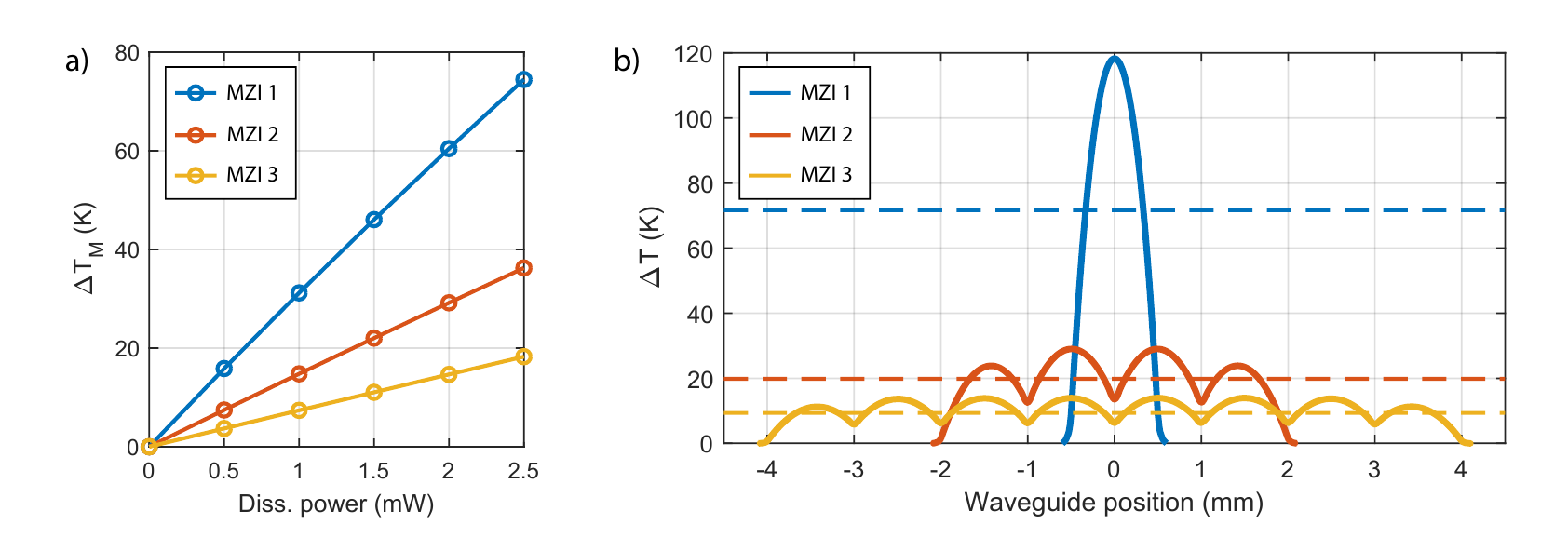}
    \caption{(a) Mean temperature variation $\Delta T_M $ along the bridge of the three MZIs as a function of the power dissipation $P$. Straight segments connect the simulated data points for eye guiding. (b) Temperature variation $\Delta T$ as a function of the position along the bridge of the three MZIs when the heaters are operated at $P=P^A_{2\pi}$. Dashed lines mark the mean temperature variation $\Delta T^{2\pi}_M$. The simulations in both panels consider room temperature and high vacuum starting conditions.} 
    \label{Fig4}
\end{figure}

In order to improve our understanding of the TOPS functioning, we performed a finite element numerical simulation of the temperature distribution along the TOPS length during electrical actuation at room temperature. More specifically, the heat equation was solved numerically (using Comsol) under the following assumptions: (i) we neglect longitudinal heat transport in the metal film, and the electrical power dissipation is included in simulations with suitable boundary conditions at the upper surface of the glass bridge (such assumption is justified by considerations reported in the supplementary material); (ii) the thermal conductivity $k_{glass}$ of the Eagle XG is included in the simulation along with its temperature dependence around room conditions \cite{corning2021}; (iii) we model vacuum as a perfect thermally insulating medium, and the only heat dissipation channels from the TOPS are the bridge termination facets and the reinforcement pillars; (iv) The temperature $T_0$ at the bottom of the glass chip, fixed to 294 K, is not affected by the device operation. This assumption is justified even at cryogenic temperature, since the cooling power of the cryostat is more than 100 times greater than the electrical power used for the TOPS actuation, and, during operation, no significant variation has been measured by the cold finger temperature sensor.
The results of these simulations indicate that the average temperature increase $\Delta T_M$ inside the suspended glass bridge is almost directly proportional to $P$, in agreement with the low values of $\beta$ observed experimentally, and it decreases for longer TOPS (see figure \ref{Fig4}(a)) for an equal amount of dissipated power. Then, we simulated the temperature distribution of the three MZIs in the case of $P=P^A_{2\pi}$. The results are shown in figure \ref{Fig4}(b) for a cutline corresponding to the heated waveguide of each MZI. At first, it is possible to appreciate that the temperature gradient has a non-negligible component parallel to the waveguide. This is a direct consequence of the strong insulation provided by the vacuum around the waveguide, which forces the heat to diffuse longitudinally in the bridge. This behavior is substantially different from operation in standard pressure conditions, in which temperature can be considered constant along the whole bridge \cite{ceccarelli2020low}. In addition, it is worth noting how the temperature outside the bridge rapidly drops to $T_0$ few tens of microns away from the TOPS at both the facets of the bridge and at the base of the pillars (not shown). From these simulations we computed the mean temperature variations $\Delta T_M^{2\pi}$ along a waveguide segment of length $L_m=L_u+$0.2 mm, which includes the temperature decay outside the bridges, and we used these values to estimate the glass thermo-optic coefficient as $c^*_{to} = \lambda /(L_m \Delta T_M^{2\pi})$ (being $\lambda$ the light wavelength in vacuum, see the supplementary material for more details). All these results are listed in table \ref{Tab1} and the values of $\Delta T_M^{2\pi}$ are visualized in figure \ref{Fig4}(b) as dashed horizontal lines. The accuracy of our simulation is demonstrated by the fact that the values of $c^*_{to}$ obtained for the three MZIs are similar to each other and in good agreement with literature values \cite{ceccarelli2020low}. Interestingly, the glass pillars underneath the bridge play a major role in determining both the temperature distribution and the MZI power dissipation, as evident by looking at the waviness of the temperature profile for MZI 2 and 3, with the temperature dropping at each pillar. This phenomenon is discussed in more detail in the supplementary material.

These results explain qualitatively also our experimental observations at cryogenic temperature. In particular, a stronger temperature increase for the 1 mm TOPS is in agreement with the fact that the efficiency reported in figure \ref{Fig3}(d) does not drop at low power as significantly as in the case of longer TOPSs reported in figure \ref{Fig3}(e-f). Indeed, the higher operating temperature of MZI 1 is at the origin of both a higher linearity and an efficiency $F(P)$ that is closer to the one measured at room temperature. This may be a preferable condition if one wants the most similar behaviour at the two temperature ranges. Nonetheless, the longer MZIs still show a better overall efficiency, with a lower $P^B_{2\pi}$. The higher efficiency of longer devices, particularly evident at room temperature, is explained by the higher insulation reached by the bridge despite of the presence of reinforcement pillars. More details on the reason why the length affects the efficiency are provided in the supplementary material.

\section{Conclusion}
In this study, we have successfully demonstrated the fabrication and characterization of a programmable glass FLW PIC equipped with TOPSs that are suitable for operation both at room and cryogenic temperatures. By leveraging the unique 3D microstructuring capabilities of the FLW technology, we achieved reliable PIC operation with minimal power consumption in both temperature regimes. Interestingly, our observations indicate that the drop of the glass thermo-optic coefficient at cryogenic temperatures only marginally impacts the TOPS functioning, given the rapid local heating of the TOPS during actuation. Numerical simulations further corroborate our experimental findings, and show that, in the case of suitable substrate thermalization, relevant temperature gradients remain confined within the TOPS volume, thus not affecting the operation of other integrated components placed in the TOPS proximity.

In conclusion, our work represents a step forward in the development of quantum photonic technologies, extending the applicability of the FLW platform to the realization of fully integrated quantum photonic devices encompassing building blocks operating at cryogenic temperatures like single photon sources, quantum memories and SNSPDs.

\section*{Data availability statement}
The data that support the findings of this study are available upon reasonable request from the authors.

\section*{Acknowledgment}
RO and GC acknowledge financial support from European Union NextGenerationEU (PNRR MUR project PE0000023 – NQSTI Spoke 7. This work was partially performed at PoliFAB, the micro- and nanofabrication facility of Politecnico di Milano (www.polifab.polimi.it). The authors would like to thank the PoliFAB staff for the valuable technical support. The authors would also like to thanks Mario Buffone for the valuable help in the device fabrication.

\section*{Orcid iDs}
Francesco Ceccarelli \url{https://orcid.org/0000-0001-9574-2899}\\
Jelena V. Rakonjac \url{https://orcid.org/0000-0002-0364-9767}\\
Samuele Grandi \url{https://orcid.org/0000-0002-6986-8292}\\
Hugues de Riedmatten \url{https://orcid.org/0000-0002-4418-0723}\\
Roberto Osellame \url{https://orcid.org/0000-0002-4457-9902}\\
Giacomo Corrielli \url{https://orcid.org/0000-0002-1329-8972}\\

\bibliography{bibliography.bib}

\begin{thebibliography}{38}%
\makeatletter
\providecommand \@ifxundefined [1]{%
 \@ifx{#1\undefined}
}%
\providecommand \@ifnum [1]{%
 \ifnum #1\expandafter \@firstoftwo
 \else \expandafter \@secondoftwo
 \fi
}%
\providecommand \@ifx [1]{%
 \ifx #1\expandafter \@firstoftwo
 \else \expandafter \@secondoftwo
 \fi
}%
\providecommand \natexlab [1]{#1}%
\providecommand \enquote  [1]{``#1''}%
\providecommand \bibnamefont  [1]{#1}%
\providecommand \bibfnamefont [1]{#1}%
\providecommand \citenamefont [1]{#1}%
\providecommand \href@noop [0]{\@secondoftwo}%
\providecommand \href [0]{\begingroup \@sanitize@url \@href}%
\providecommand \@href[1]{\@@startlink{#1}\@@href}%
\providecommand \@@href[1]{\endgroup#1\@@endlink}%
\providecommand \@sanitize@url [0]{\catcode `\\12\catcode `\$12\catcode `\&12\catcode `\#12\catcode `\^12\catcode `\_12\catcode `\%12\relax}%
\providecommand \@@startlink[1]{}%
\providecommand \@@endlink[0]{}%
\providecommand \url  [0]{\begingroup\@sanitize@url \@url }%
\providecommand \@url [1]{\endgroup\@href {#1}{\urlprefix }}%
\providecommand \urlprefix  [0]{URL }%
\providecommand \Eprint [0]{\href }%
\providecommand \doibase [0]{https://doi.org/}%
\providecommand \selectlanguage [0]{\@gobble}%
\providecommand \bibinfo  [0]{\@secondoftwo}%
\providecommand \bibfield  [0]{\@secondoftwo}%
\providecommand \translation [1]{[#1]}%
\providecommand \BibitemOpen [0]{}%
\providecommand \bibitemStop [0]{}%
\providecommand \bibitemNoStop [0]{.\EOS\space}%
\providecommand \EOS [0]{\spacefactor3000\relax}%
\providecommand \BibitemShut  [1]{\csname bibitem#1\endcsname}%
\let\auto@bib@innerbib\@empty
\bibitem [{\citenamefont {Havemann}\ and\ \citenamefont {Hutchby}(2001)}]{havemann2001high}%
  \BibitemOpen
  \bibfield  {author} {\bibinfo {author} {\bibfnamefont {R.~H.}\ \bibnamefont {Havemann}}\ and\ \bibinfo {author} {\bibfnamefont {J.~A.}\ \bibnamefont {Hutchby}},\ }\bibfield  {title} {\bibinfo {title} {High-performance interconnects: An integration overview},\ }\href@noop {} {\bibfield  {journal} {\bibinfo  {journal} {Proceedings of the IEEE}\ }\textbf {\bibinfo {volume} {89}},\ \bibinfo {pages} {586} (\bibinfo {year} {2001})}\BibitemShut {NoStop}%
\bibitem [{\citenamefont {Beausoleil}(2011)}]{beausoleil2011large}%
  \BibitemOpen
  \bibfield  {author} {\bibinfo {author} {\bibfnamefont {R.~G.}\ \bibnamefont {Beausoleil}},\ }\bibfield  {title} {\bibinfo {title} {Large-scale integrated photonics for high-performance interconnects},\ }\href@noop {} {\bibfield  {journal} {\bibinfo  {journal} {ACM Journal on Emerging Technologies in Computing Systems (JETC)}\ }\textbf {\bibinfo {volume} {7}},\ \bibinfo {pages} {1} (\bibinfo {year} {2011})}\BibitemShut {NoStop}%
\bibitem [{\citenamefont {Shastri}\ \emph {et~al.}(2021)\citenamefont {Shastri}, \citenamefont {Tait}, \citenamefont {Ferreira~de Lima}, \citenamefont {Pernice}, \citenamefont {Bhaskaran}, \citenamefont {Wright},\ and\ \citenamefont {Prucnal}}]{shastri2021photonics}%
  \BibitemOpen
  \bibfield  {author} {\bibinfo {author} {\bibfnamefont {B.~J.}\ \bibnamefont {Shastri}}, \bibinfo {author} {\bibfnamefont {A.~N.}\ \bibnamefont {Tait}}, \bibinfo {author} {\bibfnamefont {T.}~\bibnamefont {Ferreira~de Lima}}, \bibinfo {author} {\bibfnamefont {W.~H.}\ \bibnamefont {Pernice}}, \bibinfo {author} {\bibfnamefont {H.}~\bibnamefont {Bhaskaran}}, \bibinfo {author} {\bibfnamefont {C.~D.}\ \bibnamefont {Wright}},\ and\ \bibinfo {author} {\bibfnamefont {P.~R.}\ \bibnamefont {Prucnal}},\ }\bibfield  {title} {\bibinfo {title} {Photonics for artificial intelligence and neuromorphic computing},\ }\href@noop {} {\bibfield  {journal} {\bibinfo  {journal} {Nature Photonics}\ }\textbf {\bibinfo {volume} {15}},\ \bibinfo {pages} {102} (\bibinfo {year} {2021})}\BibitemShut {NoStop}%
\bibitem [{\citenamefont {Wang}\ \emph {et~al.}(2020)\citenamefont {Wang}, \citenamefont {Sciarrino}, \citenamefont {Laing},\ and\ \citenamefont {Thompson}}]{wang2020integrated}%
  \BibitemOpen
  \bibfield  {author} {\bibinfo {author} {\bibfnamefont {J.}~\bibnamefont {Wang}}, \bibinfo {author} {\bibfnamefont {F.}~\bibnamefont {Sciarrino}}, \bibinfo {author} {\bibfnamefont {A.}~\bibnamefont {Laing}},\ and\ \bibinfo {author} {\bibfnamefont {M.~G.}\ \bibnamefont {Thompson}},\ }\bibfield  {title} {\bibinfo {title} {Integrated photonic quantum technologies},\ }\href@noop {} {\bibfield  {journal} {\bibinfo  {journal} {Nature Photonics}\ }\textbf {\bibinfo {volume} {14}},\ \bibinfo {pages} {273} (\bibinfo {year} {2020})}\BibitemShut {NoStop}%
\bibitem [{\citenamefont {Davanco}\ \emph {et~al.}(2017)\citenamefont {Davanco}, \citenamefont {Liu}, \citenamefont {Sapienza}, \citenamefont {Zhang}, \citenamefont {De~Miranda~Cardoso}, \citenamefont {Verma}, \citenamefont {Mirin}, \citenamefont {Nam}, \citenamefont {Liu},\ and\ \citenamefont {Srinivasan}}]{davanco2017heterogeneous}%
  \BibitemOpen
  \bibfield  {author} {\bibinfo {author} {\bibfnamefont {M.}~\bibnamefont {Davanco}}, \bibinfo {author} {\bibfnamefont {J.}~\bibnamefont {Liu}}, \bibinfo {author} {\bibfnamefont {L.}~\bibnamefont {Sapienza}}, \bibinfo {author} {\bibfnamefont {C.-Z.}\ \bibnamefont {Zhang}}, \bibinfo {author} {\bibfnamefont {J.~V.}\ \bibnamefont {De~Miranda~Cardoso}}, \bibinfo {author} {\bibfnamefont {V.}~\bibnamefont {Verma}}, \bibinfo {author} {\bibfnamefont {R.}~\bibnamefont {Mirin}}, \bibinfo {author} {\bibfnamefont {S.~W.}\ \bibnamefont {Nam}}, \bibinfo {author} {\bibfnamefont {L.}~\bibnamefont {Liu}},\ and\ \bibinfo {author} {\bibfnamefont {K.}~\bibnamefont {Srinivasan}},\ }\bibfield  {title} {\bibinfo {title} {Heterogeneous integration for on-chip quantum photonic circuits with single quantum dot devices},\ }\href@noop {} {\bibfield  {journal} {\bibinfo  {journal} {Nature Communications}\ }\textbf {\bibinfo {volume} {8}},\ \bibinfo {pages} {889} (\bibinfo {year} {2017})}\BibitemShut {NoStop}%
\bibitem [{\citenamefont {Zhou}\ \emph {et~al.}(2023)\citenamefont {Zhou}, \citenamefont {Liu}, \citenamefont {Li}, \citenamefont {Guo}, \citenamefont {Oblak}, \citenamefont {Lei}, \citenamefont {Faraon}, \citenamefont {Mazzera},\ and\ \citenamefont {de~Riedmatten}}]{zhou2023photonic}%
  \BibitemOpen
  \bibfield  {author} {\bibinfo {author} {\bibfnamefont {Z.-Q.}\ \bibnamefont {Zhou}}, \bibinfo {author} {\bibfnamefont {C.}~\bibnamefont {Liu}}, \bibinfo {author} {\bibfnamefont {C.-F.}\ \bibnamefont {Li}}, \bibinfo {author} {\bibfnamefont {G.-C.}\ \bibnamefont {Guo}}, \bibinfo {author} {\bibfnamefont {D.}~\bibnamefont {Oblak}}, \bibinfo {author} {\bibfnamefont {M.}~\bibnamefont {Lei}}, \bibinfo {author} {\bibfnamefont {A.}~\bibnamefont {Faraon}}, \bibinfo {author} {\bibfnamefont {M.}~\bibnamefont {Mazzera}},\ and\ \bibinfo {author} {\bibfnamefont {H.}~\bibnamefont {de~Riedmatten}},\ }\bibfield  {title} {\bibinfo {title} {Photonic integrated quantum memory in rare-earth doped solids},\ }\href@noop {} {\bibfield  {journal} {\bibinfo  {journal} {Laser \& Photonics Reviews}\ }\textbf {\bibinfo {volume} {17}},\ \bibinfo {pages} {2300257} (\bibinfo {year} {2023})}\BibitemShut {NoStop}%
\bibitem [{\citenamefont {Xiong}\ \emph {et~al.}(2023)\citenamefont {Xiong}, \citenamefont {Bourgois}, \citenamefont {Sheremetyeva}, \citenamefont {Chen}, \citenamefont {Dahliah}, \citenamefont {Song}, \citenamefont {Zheng}, \citenamefont {Griffin}, \citenamefont {Sipahigil},\ and\ \citenamefont {Hautier}}]{xiong2023high}%
  \BibitemOpen
  \bibfield  {author} {\bibinfo {author} {\bibfnamefont {Y.}~\bibnamefont {Xiong}}, \bibinfo {author} {\bibfnamefont {C.}~\bibnamefont {Bourgois}}, \bibinfo {author} {\bibfnamefont {N.}~\bibnamefont {Sheremetyeva}}, \bibinfo {author} {\bibfnamefont {W.}~\bibnamefont {Chen}}, \bibinfo {author} {\bibfnamefont {D.}~\bibnamefont {Dahliah}}, \bibinfo {author} {\bibfnamefont {H.}~\bibnamefont {Song}}, \bibinfo {author} {\bibfnamefont {J.}~\bibnamefont {Zheng}}, \bibinfo {author} {\bibfnamefont {S.~M.}\ \bibnamefont {Griffin}}, \bibinfo {author} {\bibfnamefont {A.}~\bibnamefont {Sipahigil}},\ and\ \bibinfo {author} {\bibfnamefont {G.}~\bibnamefont {Hautier}},\ }\bibfield  {title} {\bibinfo {title} {High-throughput identification of spin-photon interfaces in silicon},\ }\href@noop {} {\bibfield  {journal} {\bibinfo  {journal} {Science Advances}\ }\textbf {\bibinfo {volume} {9}},\ \bibinfo {pages} {eadh8617} (\bibinfo {year} {2023})}\BibitemShut {NoStop}%
\bibitem [{\citenamefont {Esmaeil~Zadeh}\ \emph {et~al.}(2021)\citenamefont {Esmaeil~Zadeh}, \citenamefont {Chang}, \citenamefont {Los}, \citenamefont {Gyger}, \citenamefont {Elshaari}, \citenamefont {Steinhauer}, \citenamefont {Dorenbos},\ and\ \citenamefont {Zwiller}}]{esmaeil2021superconducting}%
  \BibitemOpen
  \bibfield  {author} {\bibinfo {author} {\bibfnamefont {I.}~\bibnamefont {Esmaeil~Zadeh}}, \bibinfo {author} {\bibfnamefont {J.}~\bibnamefont {Chang}}, \bibinfo {author} {\bibfnamefont {J.~W.}\ \bibnamefont {Los}}, \bibinfo {author} {\bibfnamefont {S.}~\bibnamefont {Gyger}}, \bibinfo {author} {\bibfnamefont {A.~W.}\ \bibnamefont {Elshaari}}, \bibinfo {author} {\bibfnamefont {S.}~\bibnamefont {Steinhauer}}, \bibinfo {author} {\bibfnamefont {S.~N.}\ \bibnamefont {Dorenbos}},\ and\ \bibinfo {author} {\bibfnamefont {V.}~\bibnamefont {Zwiller}},\ }\bibfield  {title} {\bibinfo {title} {Superconducting nanowire single-photon detectors: A perspective on evolution, state-of-the-art, future developments, and applications},\ }\href@noop {} {\bibfield  {journal} {\bibinfo  {journal} {Applied Physics Letters}\ }\textbf {\bibinfo {volume} {118}} (\bibinfo {year} {2021})}\BibitemShut {NoStop}%
\bibitem [{\citenamefont {Mehta}\ \emph {et~al.}(2016)\citenamefont {Mehta}, \citenamefont {Bruzewicz}, \citenamefont {McConnell}, \citenamefont {Ram}, \citenamefont {Sage},\ and\ \citenamefont {Chiaverini}}]{mehta2016integrated}%
  \BibitemOpen
  \bibfield  {author} {\bibinfo {author} {\bibfnamefont {K.~K.}\ \bibnamefont {Mehta}}, \bibinfo {author} {\bibfnamefont {C.~D.}\ \bibnamefont {Bruzewicz}}, \bibinfo {author} {\bibfnamefont {R.}~\bibnamefont {McConnell}}, \bibinfo {author} {\bibfnamefont {R.~J.}\ \bibnamefont {Ram}}, \bibinfo {author} {\bibfnamefont {J.~M.}\ \bibnamefont {Sage}},\ and\ \bibinfo {author} {\bibfnamefont {J.}~\bibnamefont {Chiaverini}},\ }\bibfield  {title} {\bibinfo {title} {Integrated optical addressing of an ion qubit},\ }\href@noop {} {\bibfield  {journal} {\bibinfo  {journal} {Nature nanotechnology}\ }\textbf {\bibinfo {volume} {11}},\ \bibinfo {pages} {1066} (\bibinfo {year} {2016})}\BibitemShut {NoStop}%
\bibitem [{\citenamefont {Timpu}\ \emph {et~al.}(2022)\citenamefont {Timpu}, \citenamefont {Matt}, \citenamefont {Piacentini}, \citenamefont {Corbelli}, \citenamefont {Marinelli}, \citenamefont {Hempel}, \citenamefont {Osellame},\ and\ \citenamefont {Home}}]{timpu2022laser}%
  \BibitemOpen
  \bibfield  {author} {\bibinfo {author} {\bibfnamefont {F.}~\bibnamefont {Timpu}}, \bibinfo {author} {\bibfnamefont {R.}~\bibnamefont {Matt}}, \bibinfo {author} {\bibfnamefont {S.}~\bibnamefont {Piacentini}}, \bibinfo {author} {\bibfnamefont {G.}~\bibnamefont {Corbelli}}, \bibinfo {author} {\bibfnamefont {M.}~\bibnamefont {Marinelli}}, \bibinfo {author} {\bibfnamefont {C.}~\bibnamefont {Hempel}}, \bibinfo {author} {\bibfnamefont {R.}~\bibnamefont {Osellame}},\ and\ \bibinfo {author} {\bibfnamefont {J.}~\bibnamefont {Home}},\ }\bibfield  {title} {\bibinfo {title} {Laser-written waveguide array optimized for individual control of trapped ion qubits in a chain},\ }in\ \href@noop {} {\emph {\bibinfo {booktitle} {European Conference and Exhibition on Optical Communication}}}\ (\bibinfo {organization} {Optica Publishing Group},\ \bibinfo {year} {2022})\ pp.\ \bibinfo {pages} {We5--70}\BibitemShut {NoStop}%
\bibitem [{\citenamefont {Lecocq}\ \emph {et~al.}(2021)\citenamefont {Lecocq}, \citenamefont {Quinlan}, \citenamefont {Cicak}, \citenamefont {Aumentado}, \citenamefont {Diddams},\ and\ \citenamefont {Teufel}}]{lecocq2021control}%
  \BibitemOpen
  \bibfield  {author} {\bibinfo {author} {\bibfnamefont {F.}~\bibnamefont {Lecocq}}, \bibinfo {author} {\bibfnamefont {F.}~\bibnamefont {Quinlan}}, \bibinfo {author} {\bibfnamefont {K.}~\bibnamefont {Cicak}}, \bibinfo {author} {\bibfnamefont {J.}~\bibnamefont {Aumentado}}, \bibinfo {author} {\bibfnamefont {S.}~\bibnamefont {Diddams}},\ and\ \bibinfo {author} {\bibfnamefont {J.}~\bibnamefont {Teufel}},\ }\bibfield  {title} {\bibinfo {title} {Control and readout of a superconducting qubit using a photonic link},\ }\href@noop {} {\bibfield  {journal} {\bibinfo  {journal} {Nature}\ }\textbf {\bibinfo {volume} {591}},\ \bibinfo {pages} {575} (\bibinfo {year} {2021})}\BibitemShut {NoStop}%
\bibitem [{\citenamefont {Pintus}\ \emph {et~al.}(2019)\citenamefont {Pintus}, \citenamefont {Zhang}, \citenamefont {Pinna}, \citenamefont {Tran}, \citenamefont {Jain}, \citenamefont {Kennedy}, \citenamefont {Ranzani}, \citenamefont {Soltani},\ and\ \citenamefont {Bowers}}]{pintus2019characterization}%
  \BibitemOpen
  \bibfield  {author} {\bibinfo {author} {\bibfnamefont {P.}~\bibnamefont {Pintus}}, \bibinfo {author} {\bibfnamefont {Z.}~\bibnamefont {Zhang}}, \bibinfo {author} {\bibfnamefont {S.}~\bibnamefont {Pinna}}, \bibinfo {author} {\bibfnamefont {M.~A.}\ \bibnamefont {Tran}}, \bibinfo {author} {\bibfnamefont {A.}~\bibnamefont {Jain}}, \bibinfo {author} {\bibfnamefont {M.}~\bibnamefont {Kennedy}}, \bibinfo {author} {\bibfnamefont {L.}~\bibnamefont {Ranzani}}, \bibinfo {author} {\bibfnamefont {M.}~\bibnamefont {Soltani}},\ and\ \bibinfo {author} {\bibfnamefont {J.~E.}\ \bibnamefont {Bowers}},\ }\bibfield  {title} {\bibinfo {title} {Characterization of heterogeneous {InP-on-Si} optical modulators operating between 77 {K} and room temperature},\ }\href@noop {} {\bibfield  {journal} {\bibinfo  {journal} {APL Photonics}\ }\textbf {\bibinfo {volume} {4}} (\bibinfo {year} {2019})}\BibitemShut {NoStop}%
\bibitem [{\citenamefont {Eltes}\ \emph {et~al.}(2020)\citenamefont {Eltes}, \citenamefont {Villarreal-Garcia}, \citenamefont {Caimi}, \citenamefont {Siegwart}, \citenamefont {Gentile}, \citenamefont {Hart}, \citenamefont {Stark}, \citenamefont {Marshall}, \citenamefont {Thompson}, \citenamefont {Barreto} \emph {et~al.}}]{eltes2020integrated}%
  \BibitemOpen
  \bibfield  {author} {\bibinfo {author} {\bibfnamefont {F.}~\bibnamefont {Eltes}}, \bibinfo {author} {\bibfnamefont {G.~E.}\ \bibnamefont {Villarreal-Garcia}}, \bibinfo {author} {\bibfnamefont {D.}~\bibnamefont {Caimi}}, \bibinfo {author} {\bibfnamefont {H.}~\bibnamefont {Siegwart}}, \bibinfo {author} {\bibfnamefont {A.~A.}\ \bibnamefont {Gentile}}, \bibinfo {author} {\bibfnamefont {A.}~\bibnamefont {Hart}}, \bibinfo {author} {\bibfnamefont {P.}~\bibnamefont {Stark}}, \bibinfo {author} {\bibfnamefont {G.~D.}\ \bibnamefont {Marshall}}, \bibinfo {author} {\bibfnamefont {M.~G.}\ \bibnamefont {Thompson}}, \bibinfo {author} {\bibfnamefont {J.}~\bibnamefont {Barreto}}, \emph {et~al.},\ }\bibfield  {title} {\bibinfo {title} {An integrated optical modulator operating at cryogenic temperatures},\ }\href@noop {} {\bibfield  {journal} {\bibinfo  {journal} {Nature Materials}\ }\textbf {\bibinfo {volume} {19}},\ \bibinfo {pages} {1164} (\bibinfo {year} {2020})}\BibitemShut {NoStop}%
\bibitem [{\citenamefont {Chakraborty}\ \emph {et~al.}(2020)\citenamefont {Chakraborty}, \citenamefont {Carolan}, \citenamefont {Clark}, \citenamefont {Bunandar}, \citenamefont {Gilbert}, \citenamefont {Notaros}, \citenamefont {Watts},\ and\ \citenamefont {Englund}}]{chakraborty2020cryogenic}%
  \BibitemOpen
  \bibfield  {author} {\bibinfo {author} {\bibfnamefont {U.}~\bibnamefont {Chakraborty}}, \bibinfo {author} {\bibfnamefont {J.}~\bibnamefont {Carolan}}, \bibinfo {author} {\bibfnamefont {G.}~\bibnamefont {Clark}}, \bibinfo {author} {\bibfnamefont {D.}~\bibnamefont {Bunandar}}, \bibinfo {author} {\bibfnamefont {G.}~\bibnamefont {Gilbert}}, \bibinfo {author} {\bibfnamefont {J.}~\bibnamefont {Notaros}}, \bibinfo {author} {\bibfnamefont {M.~R.}\ \bibnamefont {Watts}},\ and\ \bibinfo {author} {\bibfnamefont {D.~R.}\ \bibnamefont {Englund}},\ }\bibfield  {title} {\bibinfo {title} {Cryogenic operation of silicon photonic modulators based on the {DC Kerr} effect},\ }\href@noop {} {\bibfield  {journal} {\bibinfo  {journal} {Optica}\ }\textbf {\bibinfo {volume} {7}},\ \bibinfo {pages} {1385} (\bibinfo {year} {2020})}\BibitemShut {NoStop}%
\bibitem [{\citenamefont {Lee}\ \emph {et~al.}(2020)\citenamefont {Lee}, \citenamefont {Kim}, \citenamefont {Freitas}, \citenamefont {Mohanty}, \citenamefont {Zhu}, \citenamefont {Bhatt}, \citenamefont {Hone},\ and\ \citenamefont {Lipson}}]{lee2020high}%
  \BibitemOpen
  \bibfield  {author} {\bibinfo {author} {\bibfnamefont {B.~S.}\ \bibnamefont {Lee}}, \bibinfo {author} {\bibfnamefont {B.}~\bibnamefont {Kim}}, \bibinfo {author} {\bibfnamefont {A.~P.}\ \bibnamefont {Freitas}}, \bibinfo {author} {\bibfnamefont {A.}~\bibnamefont {Mohanty}}, \bibinfo {author} {\bibfnamefont {Y.}~\bibnamefont {Zhu}}, \bibinfo {author} {\bibfnamefont {G.~R.}\ \bibnamefont {Bhatt}}, \bibinfo {author} {\bibfnamefont {J.}~\bibnamefont {Hone}},\ and\ \bibinfo {author} {\bibfnamefont {M.}~\bibnamefont {Lipson}},\ }\bibfield  {title} {\bibinfo {title} {High-performance integrated graphene electro-optic modulator at cryogenic temperature},\ }\href@noop {} {\bibfield  {journal} {\bibinfo  {journal} {Nanophotonics}\ }\textbf {\bibinfo {volume} {10}},\ \bibinfo {pages} {99} (\bibinfo {year} {2020})}\BibitemShut {NoStop}%
\bibitem [{\citenamefont {Youssefi}\ \emph {et~al.}(2021)\citenamefont {Youssefi}, \citenamefont {Shomroni}, \citenamefont {Joshi}, \citenamefont {Bernier}, \citenamefont {Lukashchuk}, \citenamefont {Uhrich}, \citenamefont {Qiu},\ and\ \citenamefont {Kippenberg}}]{youssefi2021cryogenic}%
  \BibitemOpen
  \bibfield  {author} {\bibinfo {author} {\bibfnamefont {A.}~\bibnamefont {Youssefi}}, \bibinfo {author} {\bibfnamefont {I.}~\bibnamefont {Shomroni}}, \bibinfo {author} {\bibfnamefont {Y.~J.}\ \bibnamefont {Joshi}}, \bibinfo {author} {\bibfnamefont {N.~R.}\ \bibnamefont {Bernier}}, \bibinfo {author} {\bibfnamefont {A.}~\bibnamefont {Lukashchuk}}, \bibinfo {author} {\bibfnamefont {P.}~\bibnamefont {Uhrich}}, \bibinfo {author} {\bibfnamefont {L.}~\bibnamefont {Qiu}},\ and\ \bibinfo {author} {\bibfnamefont {T.~J.}\ \bibnamefont {Kippenberg}},\ }\bibfield  {title} {\bibinfo {title} {A cryogenic electro-optic interconnect for superconducting devices},\ }\href@noop {} {\bibfield  {journal} {\bibinfo  {journal} {Nature Electronics}\ }\textbf {\bibinfo {volume} {4}},\ \bibinfo {pages} {326} (\bibinfo {year} {2021})}\BibitemShut {NoStop}%
\bibitem [{\citenamefont {Lomonte}\ \emph {et~al.}(2021)\citenamefont {Lomonte}, \citenamefont {Wolff}, \citenamefont {Beutel}, \citenamefont {Ferrari}, \citenamefont {Schuck}, \citenamefont {Pernice},\ and\ \citenamefont {Lenzini}}]{lomonte2021single}%
  \BibitemOpen
  \bibfield  {author} {\bibinfo {author} {\bibfnamefont {E.}~\bibnamefont {Lomonte}}, \bibinfo {author} {\bibfnamefont {M.~A.}\ \bibnamefont {Wolff}}, \bibinfo {author} {\bibfnamefont {F.}~\bibnamefont {Beutel}}, \bibinfo {author} {\bibfnamefont {S.}~\bibnamefont {Ferrari}}, \bibinfo {author} {\bibfnamefont {C.}~\bibnamefont {Schuck}}, \bibinfo {author} {\bibfnamefont {W.~H.}\ \bibnamefont {Pernice}},\ and\ \bibinfo {author} {\bibfnamefont {F.}~\bibnamefont {Lenzini}},\ }\bibfield  {title} {\bibinfo {title} {Single-photon detection and cryogenic reconfigurability in lithium niobate nanophotonic circuits},\ }\href@noop {} {\bibfield  {journal} {\bibinfo  {journal} {Nature Communications}\ }\textbf {\bibinfo {volume} {12}},\ \bibinfo {pages} {6847} (\bibinfo {year} {2021})}\BibitemShut {NoStop}%
\bibitem [{\citenamefont {Gehl}\ \emph {et~al.}(2017)\citenamefont {Gehl}, \citenamefont {Long}, \citenamefont {Trotter}, \citenamefont {Starbuck}, \citenamefont {Pomerene}, \citenamefont {Wright}, \citenamefont {Melgaard}, \citenamefont {Siirola}, \citenamefont {Lentine},\ and\ \citenamefont {DeRose}}]{gehl2017operation}%
  \BibitemOpen
  \bibfield  {author} {\bibinfo {author} {\bibfnamefont {M.}~\bibnamefont {Gehl}}, \bibinfo {author} {\bibfnamefont {C.}~\bibnamefont {Long}}, \bibinfo {author} {\bibfnamefont {D.}~\bibnamefont {Trotter}}, \bibinfo {author} {\bibfnamefont {A.}~\bibnamefont {Starbuck}}, \bibinfo {author} {\bibfnamefont {A.}~\bibnamefont {Pomerene}}, \bibinfo {author} {\bibfnamefont {J.~B.}\ \bibnamefont {Wright}}, \bibinfo {author} {\bibfnamefont {S.}~\bibnamefont {Melgaard}}, \bibinfo {author} {\bibfnamefont {J.}~\bibnamefont {Siirola}}, \bibinfo {author} {\bibfnamefont {A.~L.}\ \bibnamefont {Lentine}},\ and\ \bibinfo {author} {\bibfnamefont {C.}~\bibnamefont {DeRose}},\ }\bibfield  {title} {\bibinfo {title} {Operation of high-speed silicon photonic micro-disk modulators at cryogenic temperatures},\ }\href@noop {} {\bibfield  {journal} {\bibinfo  {journal} {Optica}\ }\textbf {\bibinfo {volume} {4}},\ \bibinfo {pages} {374} (\bibinfo {year} {2017})}\BibitemShut {NoStop}%
\bibitem [{\citenamefont {Pintus}\ \emph {et~al.}(2022)\citenamefont {Pintus}, \citenamefont {Ranzani}, \citenamefont {Pinna}, \citenamefont {Huang}, \citenamefont {Gustafsson}, \citenamefont {Karinou}, \citenamefont {Casula}, \citenamefont {Shoji}, \citenamefont {Takamura}, \citenamefont {Mizumoto} \emph {et~al.}}]{pintus2022integrated}%
  \BibitemOpen
  \bibfield  {author} {\bibinfo {author} {\bibfnamefont {P.}~\bibnamefont {Pintus}}, \bibinfo {author} {\bibfnamefont {L.}~\bibnamefont {Ranzani}}, \bibinfo {author} {\bibfnamefont {S.}~\bibnamefont {Pinna}}, \bibinfo {author} {\bibfnamefont {D.}~\bibnamefont {Huang}}, \bibinfo {author} {\bibfnamefont {M.~V.}\ \bibnamefont {Gustafsson}}, \bibinfo {author} {\bibfnamefont {F.}~\bibnamefont {Karinou}}, \bibinfo {author} {\bibfnamefont {G.~A.}\ \bibnamefont {Casula}}, \bibinfo {author} {\bibfnamefont {Y.}~\bibnamefont {Shoji}}, \bibinfo {author} {\bibfnamefont {Y.}~\bibnamefont {Takamura}}, \bibinfo {author} {\bibfnamefont {T.}~\bibnamefont {Mizumoto}}, \emph {et~al.},\ }\bibfield  {title} {\bibinfo {title} {An integrated magneto-optic modulator for cryogenic applications},\ }\href@noop {} {\bibfield  {journal} {\bibinfo  {journal} {Nature Electronics}\ }\textbf {\bibinfo {volume} {5}},\ \bibinfo {pages} {604} (\bibinfo {year} {2022})}\BibitemShut {NoStop}%
\bibitem [{\citenamefont {Dong}\ \emph {et~al.}(2021)\citenamefont {Dong}, \citenamefont {Clark}, \citenamefont {Leenheer}, \citenamefont {Zimmermann}, \citenamefont {Dominguez}, \citenamefont {Menssen}, \citenamefont {Heim}, \citenamefont {Gilbert}, \citenamefont {Englund},\ and\ \citenamefont {Eichenfield}}]{dong2021cryogenically}%
  \BibitemOpen
  \bibfield  {author} {\bibinfo {author} {\bibfnamefont {M.}~\bibnamefont {Dong}}, \bibinfo {author} {\bibfnamefont {G.}~\bibnamefont {Clark}}, \bibinfo {author} {\bibfnamefont {A.}~\bibnamefont {Leenheer}}, \bibinfo {author} {\bibfnamefont {M.}~\bibnamefont {Zimmermann}}, \bibinfo {author} {\bibfnamefont {D.}~\bibnamefont {Dominguez}}, \bibinfo {author} {\bibfnamefont {A.}~\bibnamefont {Menssen}}, \bibinfo {author} {\bibfnamefont {D.}~\bibnamefont {Heim}}, \bibinfo {author} {\bibfnamefont {G.}~\bibnamefont {Gilbert}}, \bibinfo {author} {\bibfnamefont {D.}~\bibnamefont {Englund}},\ and\ \bibinfo {author} {\bibfnamefont {M.}~\bibnamefont {Eichenfield}},\ }\bibfield  {title} {\bibinfo {title} {High-speed programmable photonic circuits in a cryogenically compatible, visible-{NIR} 200 mm {CMOS} architecture},\ }\href@noop {} {\bibfield  {journal} {\bibinfo  {journal} {Nature Photonics}\ }\textbf {\bibinfo {volume} {16}},\ \bibinfo {pages} {59} (\bibinfo {year} {2021})}\BibitemShut {NoStop}%
\bibitem [{\citenamefont {Beutel}\ \emph {et~al.}(2022)\citenamefont {Beutel}, \citenamefont {Grottke}, \citenamefont {Wolff}, \citenamefont {Schuck},\ and\ \citenamefont {Pernice}}]{beutel2022cryo}%
  \BibitemOpen
  \bibfield  {author} {\bibinfo {author} {\bibfnamefont {F.}~\bibnamefont {Beutel}}, \bibinfo {author} {\bibfnamefont {T.}~\bibnamefont {Grottke}}, \bibinfo {author} {\bibfnamefont {M.~A.}\ \bibnamefont {Wolff}}, \bibinfo {author} {\bibfnamefont {C.}~\bibnamefont {Schuck}},\ and\ \bibinfo {author} {\bibfnamefont {W.~H.}\ \bibnamefont {Pernice}},\ }\bibfield  {title} {\bibinfo {title} {Cryo-compatible opto-mechanical low-voltage phase-modulator integrated with superconducting single-photon detectors},\ }\href@noop {} {\bibfield  {journal} {\bibinfo  {journal} {Optics Express}\ }\textbf {\bibinfo {volume} {30}},\ \bibinfo {pages} {30066} (\bibinfo {year} {2022})}\BibitemShut {NoStop}%
\bibitem [{\citenamefont {Gyger}\ \emph {et~al.}(2021)\citenamefont {Gyger}, \citenamefont {Zichi}, \citenamefont {Schweickert}, \citenamefont {Elshaari}, \citenamefont {Steinhauer}, \citenamefont {Covre~da Silva}, \citenamefont {Rastelli}, \citenamefont {Zwiller}, \citenamefont {J{\"o}ns},\ and\ \citenamefont {Errando-Herranz}}]{gyger2021reconfigurable}%
  \BibitemOpen
  \bibfield  {author} {\bibinfo {author} {\bibfnamefont {S.}~\bibnamefont {Gyger}}, \bibinfo {author} {\bibfnamefont {J.}~\bibnamefont {Zichi}}, \bibinfo {author} {\bibfnamefont {L.}~\bibnamefont {Schweickert}}, \bibinfo {author} {\bibfnamefont {A.~W.}\ \bibnamefont {Elshaari}}, \bibinfo {author} {\bibfnamefont {S.}~\bibnamefont {Steinhauer}}, \bibinfo {author} {\bibfnamefont {S.~F.}\ \bibnamefont {Covre~da Silva}}, \bibinfo {author} {\bibfnamefont {A.}~\bibnamefont {Rastelli}}, \bibinfo {author} {\bibfnamefont {V.}~\bibnamefont {Zwiller}}, \bibinfo {author} {\bibfnamefont {K.~D.}\ \bibnamefont {J{\"o}ns}},\ and\ \bibinfo {author} {\bibfnamefont {C.}~\bibnamefont {Errando-Herranz}},\ }\bibfield  {title} {\bibinfo {title} {Reconfigurable photonics with on-chip single-photon detectors},\ }\href@noop {} {\bibfield  {journal} {\bibinfo  {journal} {Nature Communications}\ }\textbf {\bibinfo {volume} {12}},\ \bibinfo {pages} {1408} (\bibinfo {year} {2021})}\BibitemShut {NoStop}%
\bibitem [{\citenamefont {Harris}\ \emph {et~al.}(2014)\citenamefont {Harris}, \citenamefont {Ma}, \citenamefont {Mower}, \citenamefont {Baehr-Jones}, \citenamefont {Englund}, \citenamefont {Hochberg},\ and\ \citenamefont {Galland}}]{harris2014efficient}%
  \BibitemOpen
  \bibfield  {author} {\bibinfo {author} {\bibfnamefont {N.~C.}\ \bibnamefont {Harris}}, \bibinfo {author} {\bibfnamefont {Y.}~\bibnamefont {Ma}}, \bibinfo {author} {\bibfnamefont {J.}~\bibnamefont {Mower}}, \bibinfo {author} {\bibfnamefont {T.}~\bibnamefont {Baehr-Jones}}, \bibinfo {author} {\bibfnamefont {D.}~\bibnamefont {Englund}}, \bibinfo {author} {\bibfnamefont {M.}~\bibnamefont {Hochberg}},\ and\ \bibinfo {author} {\bibfnamefont {C.}~\bibnamefont {Galland}},\ }\bibfield  {title} {\bibinfo {title} {Efficient, compact and low loss thermo-optic phase shifter in silicon},\ }\href@noop {} {\bibfield  {journal} {\bibinfo  {journal} {Optics Express}\ }\textbf {\bibinfo {volume} {22}},\ \bibinfo {pages} {10487} (\bibinfo {year} {2014})}\BibitemShut {NoStop}%
\bibitem [{\citenamefont {Tong}\ \emph {et~al.}(2023)\citenamefont {Tong}, \citenamefont {Yang}, \citenamefont {Pang}, \citenamefont {Yang}, \citenamefont {Qian}, \citenamefont {Yang}, \citenamefont {Hu}, \citenamefont {Dong},\ and\ \citenamefont {Zhang}}]{tong2023efficient}%
  \BibitemOpen
  \bibfield  {author} {\bibinfo {author} {\bibfnamefont {W.}~\bibnamefont {Tong}}, \bibinfo {author} {\bibfnamefont {E.}~\bibnamefont {Yang}}, \bibinfo {author} {\bibfnamefont {Y.}~\bibnamefont {Pang}}, \bibinfo {author} {\bibfnamefont {H.}~\bibnamefont {Yang}}, \bibinfo {author} {\bibfnamefont {X.}~\bibnamefont {Qian}}, \bibinfo {author} {\bibfnamefont {R.}~\bibnamefont {Yang}}, \bibinfo {author} {\bibfnamefont {B.}~\bibnamefont {Hu}}, \bibinfo {author} {\bibfnamefont {J.}~\bibnamefont {Dong}},\ and\ \bibinfo {author} {\bibfnamefont {X.}~\bibnamefont {Zhang}},\ }\bibfield  {title} {\bibinfo {title} {An efficient, fast-responding, low-loss thermo-optic phase shifter based on a hydrogen-doped indium oxide microheater},\ }\href@noop {} {\bibfield  {journal} {\bibinfo  {journal} {Laser \& Photonics Reviews}\ }\textbf {\bibinfo {volume} {17}},\ \bibinfo {pages} {2201032} (\bibinfo {year} {2023})}\BibitemShut {NoStop}%
\bibitem [{\citenamefont {Suzuki}\ \emph {et~al.}(2019)\citenamefont {Suzuki}, \citenamefont {Konoike}, \citenamefont {Hasegawa}, \citenamefont {Suda}, \citenamefont {Matsuura}, \citenamefont {Ikeda}, \citenamefont {Namiki},\ and\ \citenamefont {Kawashima}}]{suzuki2019low}%
  \BibitemOpen
  \bibfield  {author} {\bibinfo {author} {\bibfnamefont {K.}~\bibnamefont {Suzuki}}, \bibinfo {author} {\bibfnamefont {R.}~\bibnamefont {Konoike}}, \bibinfo {author} {\bibfnamefont {J.}~\bibnamefont {Hasegawa}}, \bibinfo {author} {\bibfnamefont {S.}~\bibnamefont {Suda}}, \bibinfo {author} {\bibfnamefont {H.}~\bibnamefont {Matsuura}}, \bibinfo {author} {\bibfnamefont {K.}~\bibnamefont {Ikeda}}, \bibinfo {author} {\bibfnamefont {S.}~\bibnamefont {Namiki}},\ and\ \bibinfo {author} {\bibfnamefont {H.}~\bibnamefont {Kawashima}},\ }\bibfield  {title} {\bibinfo {title} {Low-insertion-loss and power-efficient 32$\times$ 32 silicon photonics switch with extremely high-$\delta$ silica {PLC} connector},\ }\href@noop {} {\bibfield  {journal} {\bibinfo  {journal} {Journal of Lightwave Technology}\ }\textbf {\bibinfo {volume} {37}},\ \bibinfo {pages} {116} (\bibinfo {year} {2019})}\BibitemShut {NoStop}%
\bibitem [{\citenamefont {Komma}\ \emph {et~al.}(2012)\citenamefont {Komma}, \citenamefont {Schwarz}, \citenamefont {Hofmann}, \citenamefont {Heinert},\ and\ \citenamefont {Nawrodt}}]{komma2012thermo}%
  \BibitemOpen
  \bibfield  {author} {\bibinfo {author} {\bibfnamefont {J.}~\bibnamefont {Komma}}, \bibinfo {author} {\bibfnamefont {C.}~\bibnamefont {Schwarz}}, \bibinfo {author} {\bibfnamefont {G.}~\bibnamefont {Hofmann}}, \bibinfo {author} {\bibfnamefont {D.}~\bibnamefont {Heinert}},\ and\ \bibinfo {author} {\bibfnamefont {R.}~\bibnamefont {Nawrodt}},\ }\bibfield  {title} {\bibinfo {title} {Thermo-optic coefficient of silicon at 1550 nm and cryogenic temperatures},\ }\href@noop {} {\bibfield  {journal} {\bibinfo  {journal} {Applied Physics Letters}\ }\textbf {\bibinfo {volume} {101}} (\bibinfo {year} {2012})}\BibitemShut {NoStop}%
\bibitem [{\citenamefont {Elshaari}\ \emph {et~al.}(2016)\citenamefont {Elshaari}, \citenamefont {Zadeh}, \citenamefont {J{\"o}ns},\ and\ \citenamefont {Zwiller}}]{elshaari2016thermo}%
  \BibitemOpen
  \bibfield  {author} {\bibinfo {author} {\bibfnamefont {A.~W.}\ \bibnamefont {Elshaari}}, \bibinfo {author} {\bibfnamefont {I.~E.}\ \bibnamefont {Zadeh}}, \bibinfo {author} {\bibfnamefont {K.~D.}\ \bibnamefont {J{\"o}ns}},\ and\ \bibinfo {author} {\bibfnamefont {V.}~\bibnamefont {Zwiller}},\ }\bibfield  {title} {\bibinfo {title} {Thermo-optic characterization of silicon nitride resonators for cryogenic photonic circuits},\ }\href@noop {} {\bibfield  {journal} {\bibinfo  {journal} {IEEE Photonics Journal}\ }\textbf {\bibinfo {volume} {8}},\ \bibinfo {pages} {1} (\bibinfo {year} {2016})}\BibitemShut {NoStop}%
\bibitem [{\citenamefont {Han}\ \emph {et~al.}(2023)\citenamefont {Han}, \citenamefont {Zhang}, \citenamefont {Xiao}, \citenamefont {Yuan}, \citenamefont {Yu}, \citenamefont {Wang}, \citenamefont {Li}, \citenamefont {Xie}, \citenamefont {Lu}, \citenamefont {Li} \emph {et~al.}}]{han2023cryogenic}%
  \BibitemOpen
  \bibfield  {author} {\bibinfo {author} {\bibfnamefont {H.}~\bibnamefont {Han}}, \bibinfo {author} {\bibfnamefont {X.}~\bibnamefont {Zhang}}, \bibinfo {author} {\bibfnamefont {Y.}~\bibnamefont {Xiao}}, \bibinfo {author} {\bibfnamefont {P.}~\bibnamefont {Yuan}}, \bibinfo {author} {\bibfnamefont {H.}~\bibnamefont {Yu}}, \bibinfo {author} {\bibfnamefont {S.}~\bibnamefont {Wang}}, \bibinfo {author} {\bibfnamefont {H.}~\bibnamefont {Li}}, \bibinfo {author} {\bibfnamefont {W.}~\bibnamefont {Xie}}, \bibinfo {author} {\bibfnamefont {M.}~\bibnamefont {Lu}}, \bibinfo {author} {\bibfnamefont {L.}~\bibnamefont {Li}}, \emph {et~al.},\ }\bibfield  {title} {\bibinfo {title} {Cryogenic thermo-optic thin-film lithium niobate modulator with an {NbN} superconducting heater},\ }\href@noop {} {\bibfield  {journal} {\bibinfo  {journal} {Chinese Optics Letters}\ }\textbf {\bibinfo {volume} {21}},\ \bibinfo {pages} {081301} (\bibinfo {year} {2023})}\BibitemShut {NoStop}%
\bibitem [{\citenamefont {Ceccarelli}\ \emph {et~al.}(2020)\citenamefont {Ceccarelli}, \citenamefont {Atzeni}, \citenamefont {Pentangelo}, \citenamefont {Pellegatta}, \citenamefont {Crespi},\ and\ \citenamefont {Osellame}}]{ceccarelli2020low}%
  \BibitemOpen
  \bibfield  {author} {\bibinfo {author} {\bibfnamefont {F.}~\bibnamefont {Ceccarelli}}, \bibinfo {author} {\bibfnamefont {S.}~\bibnamefont {Atzeni}}, \bibinfo {author} {\bibfnamefont {C.}~\bibnamefont {Pentangelo}}, \bibinfo {author} {\bibfnamefont {F.}~\bibnamefont {Pellegatta}}, \bibinfo {author} {\bibfnamefont {A.}~\bibnamefont {Crespi}},\ and\ \bibinfo {author} {\bibfnamefont {R.}~\bibnamefont {Osellame}},\ }\bibfield  {title} {\bibinfo {title} {Low power reconfigurability and reduced crosstalk in integrated photonic circuits fabricated by femtosecond laser micromachining},\ }\href@noop {} {\bibfield  {journal} {\bibinfo  {journal} {Laser \& Photonics Reviews}\ }\textbf {\bibinfo {volume} {14}},\ \bibinfo {pages} {2000024} (\bibinfo {year} {2020})}\BibitemShut {NoStop}%
\bibitem [{\citenamefont {Corrielli}\ \emph {et~al.}(2021)\citenamefont {Corrielli}, \citenamefont {Crespi},\ and\ \citenamefont {Osellame}}]{corrielli2021femtosecond}%
  \BibitemOpen
  \bibfield  {author} {\bibinfo {author} {\bibfnamefont {G.}~\bibnamefont {Corrielli}}, \bibinfo {author} {\bibfnamefont {A.}~\bibnamefont {Crespi}},\ and\ \bibinfo {author} {\bibfnamefont {R.}~\bibnamefont {Osellame}},\ }\bibfield  {title} {\bibinfo {title} {Femtosecond laser micromachining for integrated quantum photonics},\ }\href@noop {} {\bibfield  {journal} {\bibinfo  {journal} {Nanophotonics}\ }\textbf {\bibinfo {volume} {10}},\ \bibinfo {pages} {3789} (\bibinfo {year} {2021})}\BibitemShut {NoStop}%
\bibitem [{\citenamefont {Dyakonov}\ \emph {et~al.}(2018)\citenamefont {Dyakonov}, \citenamefont {Pogorelov}, \citenamefont {Bobrov}, \citenamefont {Kalinkin}, \citenamefont {Straupe}, \citenamefont {Kulik}, \citenamefont {Dyakonov},\ and\ \citenamefont {Evlashin}}]{dyakonov2018reconfigurable}%
  \BibitemOpen
  \bibfield  {author} {\bibinfo {author} {\bibfnamefont {I.}~\bibnamefont {Dyakonov}}, \bibinfo {author} {\bibfnamefont {I.}~\bibnamefont {Pogorelov}}, \bibinfo {author} {\bibfnamefont {I.}~\bibnamefont {Bobrov}}, \bibinfo {author} {\bibfnamefont {A.}~\bibnamefont {Kalinkin}}, \bibinfo {author} {\bibfnamefont {S.}~\bibnamefont {Straupe}}, \bibinfo {author} {\bibfnamefont {S.}~\bibnamefont {Kulik}}, \bibinfo {author} {\bibfnamefont {P.}~\bibnamefont {Dyakonov}},\ and\ \bibinfo {author} {\bibfnamefont {S.}~\bibnamefont {Evlashin}},\ }\bibfield  {title} {\bibinfo {title} {Reconfigurable photonics on a glass chip},\ }\href@noop {} {\bibfield  {journal} {\bibinfo  {journal} {Physical Review Applied}\ }\textbf {\bibinfo {volume} {10}},\ \bibinfo {pages} {044048} (\bibinfo {year} {2018})}\BibitemShut {NoStop}%
\bibitem [{\citenamefont {Pentangelo}\ \emph {et~al.}(2024)\citenamefont {Pentangelo}, \citenamefont {Di~Giano}, \citenamefont {Piacentini}, \citenamefont {Arpe}, \citenamefont {Ceccarelli}, \citenamefont {Crespi},\ and\ \citenamefont {Osellame}}]{pentangelo2024high}%
  \BibitemOpen
  \bibfield  {author} {\bibinfo {author} {\bibfnamefont {C.}~\bibnamefont {Pentangelo}}, \bibinfo {author} {\bibfnamefont {N.}~\bibnamefont {Di~Giano}}, \bibinfo {author} {\bibfnamefont {S.}~\bibnamefont {Piacentini}}, \bibinfo {author} {\bibfnamefont {R.}~\bibnamefont {Arpe}}, \bibinfo {author} {\bibfnamefont {F.}~\bibnamefont {Ceccarelli}}, \bibinfo {author} {\bibfnamefont {A.}~\bibnamefont {Crespi}},\ and\ \bibinfo {author} {\bibfnamefont {R.}~\bibnamefont {Osellame}},\ }\bibfield  {title} {\bibinfo {title} {High-fidelity and polarization-insensitive universal photonic processors fabricated by femtosecond laser writing},\ }\href@noop {} {\bibfield  {journal} {\bibinfo  {journal} {Nanophotonics}\ } (\bibinfo {year} {2024})}\BibitemShut {NoStop}%
\bibitem [{\citenamefont {Ant{\'o}n}\ \emph {et~al.}(2019)\citenamefont {Ant{\'o}n}, \citenamefont {Loredo}, \citenamefont {Coppola}, \citenamefont {Ollivier}, \citenamefont {Viggianiello}, \citenamefont {Harouri}, \citenamefont {Somaschi}, \citenamefont {Crespi}, \citenamefont {Sagnes}, \citenamefont {Lemaitre} \emph {et~al.}}]{anton2019interfacing}%
  \BibitemOpen
  \bibfield  {author} {\bibinfo {author} {\bibfnamefont {C.}~\bibnamefont {Ant{\'o}n}}, \bibinfo {author} {\bibfnamefont {J.~C.}\ \bibnamefont {Loredo}}, \bibinfo {author} {\bibfnamefont {G.}~\bibnamefont {Coppola}}, \bibinfo {author} {\bibfnamefont {H.}~\bibnamefont {Ollivier}}, \bibinfo {author} {\bibfnamefont {N.}~\bibnamefont {Viggianiello}}, \bibinfo {author} {\bibfnamefont {A.}~\bibnamefont {Harouri}}, \bibinfo {author} {\bibfnamefont {N.}~\bibnamefont {Somaschi}}, \bibinfo {author} {\bibfnamefont {A.}~\bibnamefont {Crespi}}, \bibinfo {author} {\bibfnamefont {I.}~\bibnamefont {Sagnes}}, \bibinfo {author} {\bibfnamefont {A.}~\bibnamefont {Lemaitre}}, \emph {et~al.},\ }\bibfield  {title} {\bibinfo {title} {Interfacing scalable photonic platforms: solid-state based multi-photon interference in a reconfigurable glass chip},\ }\href@noop {} {\bibfield  {journal} {\bibinfo  {journal} {Optica}\ }\textbf {\bibinfo {volume} {6}},\ \bibinfo {pages} {1471} (\bibinfo {year} {2019})}\BibitemShut {NoStop}%
\bibitem [{\citenamefont {Rakonjac}\ \emph {et~al.}(2022)\citenamefont {Rakonjac}, \citenamefont {Corrielli}, \citenamefont {Lago-Rivera}, \citenamefont {Seri}, \citenamefont {Mazzera}, \citenamefont {Grandi}, \citenamefont {Osellame},\ and\ \citenamefont {de~Riedmatten}}]{rakonjac2022storage}%
  \BibitemOpen
  \bibfield  {author} {\bibinfo {author} {\bibfnamefont {J.~V.}\ \bibnamefont {Rakonjac}}, \bibinfo {author} {\bibfnamefont {G.}~\bibnamefont {Corrielli}}, \bibinfo {author} {\bibfnamefont {D.}~\bibnamefont {Lago-Rivera}}, \bibinfo {author} {\bibfnamefont {A.}~\bibnamefont {Seri}}, \bibinfo {author} {\bibfnamefont {M.}~\bibnamefont {Mazzera}}, \bibinfo {author} {\bibfnamefont {S.}~\bibnamefont {Grandi}}, \bibinfo {author} {\bibfnamefont {R.}~\bibnamefont {Osellame}},\ and\ \bibinfo {author} {\bibfnamefont {H.}~\bibnamefont {de~Riedmatten}},\ }\bibfield  {title} {\bibinfo {title} {Storage and analysis of light-matter entanglement in a fiber-integrated system},\ }\href@noop {} {\bibfield  {journal} {\bibinfo  {journal} {Science Advances}\ }\textbf {\bibinfo {volume} {8}},\ \bibinfo {pages} {eabn3919} (\bibinfo {year} {2022})}\BibitemShut {NoStop}%
\bibitem [{\citenamefont {Zhang}\ \emph {et~al.}(2023)\citenamefont {Zhang}, \citenamefont {Zhang}, \citenamefont {Wei}, \citenamefont {Li}, \citenamefont {Liao}, \citenamefont {Li}, \citenamefont {Deng}, \citenamefont {Wang}, \citenamefont {Song}, \citenamefont {You} \emph {et~al.}}]{zhang2023telecom}%
  \BibitemOpen
  \bibfield  {author} {\bibinfo {author} {\bibfnamefont {X.}~\bibnamefont {Zhang}}, \bibinfo {author} {\bibfnamefont {B.}~\bibnamefont {Zhang}}, \bibinfo {author} {\bibfnamefont {S.}~\bibnamefont {Wei}}, \bibinfo {author} {\bibfnamefont {H.}~\bibnamefont {Li}}, \bibinfo {author} {\bibfnamefont {J.}~\bibnamefont {Liao}}, \bibinfo {author} {\bibfnamefont {C.}~\bibnamefont {Li}}, \bibinfo {author} {\bibfnamefont {G.}~\bibnamefont {Deng}}, \bibinfo {author} {\bibfnamefont {Y.}~\bibnamefont {Wang}}, \bibinfo {author} {\bibfnamefont {H.}~\bibnamefont {Song}}, \bibinfo {author} {\bibfnamefont {L.}~\bibnamefont {You}}, \emph {et~al.},\ }\bibfield  {title} {\bibinfo {title} {Telecom-band--integrated multimode photonic quantum memory},\ }\href@noop {} {\bibfield  {journal} {\bibinfo  {journal} {Science Advances}\ }\textbf {\bibinfo {volume} {9}},\ \bibinfo {pages} {eadf4587} (\bibinfo {year} {2023})}\BibitemShut {NoStop}%
\bibitem [{\citenamefont {Koch}\ \emph {et~al.}(2022)\citenamefont {Koch}, \citenamefont {Hoese}, \citenamefont {Bharadwaj}, \citenamefont {Lang}, \citenamefont {Hadden}, \citenamefont {Ramponi}, \citenamefont {Jelezko}, \citenamefont {Eaton},\ and\ \citenamefont {Kubanek}}]{koch2022super}%
  \BibitemOpen
  \bibfield  {author} {\bibinfo {author} {\bibfnamefont {M.~K.}\ \bibnamefont {Koch}}, \bibinfo {author} {\bibfnamefont {M.}~\bibnamefont {Hoese}}, \bibinfo {author} {\bibfnamefont {V.}~\bibnamefont {Bharadwaj}}, \bibinfo {author} {\bibfnamefont {J.}~\bibnamefont {Lang}}, \bibinfo {author} {\bibfnamefont {J.~P.}\ \bibnamefont {Hadden}}, \bibinfo {author} {\bibfnamefont {R.}~\bibnamefont {Ramponi}}, \bibinfo {author} {\bibfnamefont {F.}~\bibnamefont {Jelezko}}, \bibinfo {author} {\bibfnamefont {S.~M.}\ \bibnamefont {Eaton}},\ and\ \bibinfo {author} {\bibfnamefont {A.}~\bibnamefont {Kubanek}},\ }\bibfield  {title} {\bibinfo {title} {Super-poissonian light statistics from individual silicon vacancy centers coupled to a laser-written diamond waveguide},\ }\href@noop {} {\bibfield  {journal} {\bibinfo  {journal} {ACS Photonics}\ }\textbf {\bibinfo {volume} {9}},\ \bibinfo {pages} {3366} (\bibinfo {year} {2022})}\BibitemShut {NoStop}%
\bibitem [{\citenamefont {Nayak}\ \emph {et~al.}(2021)\citenamefont {Nayak}, \citenamefont {Labadie}, \citenamefont {Sharma}, \citenamefont {Piacentini}, \citenamefont {Corrielli}, \citenamefont {Osellame}, \citenamefont {Gendron}, \citenamefont {Buey}, \citenamefont {Chemla}, \citenamefont {Cohen} \emph {et~al.}}]{nayak2021first}%
  \BibitemOpen
  \bibfield  {author} {\bibinfo {author} {\bibfnamefont {A.~S.}\ \bibnamefont {Nayak}}, \bibinfo {author} {\bibfnamefont {L.}~\bibnamefont {Labadie}}, \bibinfo {author} {\bibfnamefont {T.~K.}\ \bibnamefont {Sharma}}, \bibinfo {author} {\bibfnamefont {S.}~\bibnamefont {Piacentini}}, \bibinfo {author} {\bibfnamefont {G.}~\bibnamefont {Corrielli}}, \bibinfo {author} {\bibfnamefont {R.}~\bibnamefont {Osellame}}, \bibinfo {author} {\bibfnamefont {{\'E}.}~\bibnamefont {Gendron}}, \bibinfo {author} {\bibfnamefont {J.-T.~M.}\ \bibnamefont {Buey}}, \bibinfo {author} {\bibfnamefont {F.}~\bibnamefont {Chemla}}, \bibinfo {author} {\bibfnamefont {M.}~\bibnamefont {Cohen}}, \emph {et~al.},\ }\bibfield  {title} {\bibinfo {title} {First stellar photons for an integrated optics discrete beam combiner at the {William Herschel Telescope}},\ }\href@noop {} {\bibfield  {journal} {\bibinfo  {journal} {Applied Optics}\ }\textbf {\bibinfo {volume} {60}},\ \bibinfo {pages} {D129} (\bibinfo {year} {2021})}\BibitemShut {NoStop}%
\bibitem [{cor(2021)}]{corning2021}%
  \BibitemOpen
  \href {https://www.corning.com/media/worldwide/cdt/documents/EAGLE%20XG_PI%20Sheet_2021.pdf} {\bibinfo {title} {{Corning Eagle XG glass datasheet}}} (\bibinfo {year} {2021})\BibitemShut {NoStop}%
\end{thebibliography}%


\begin{thebibliography}{4}%
\makeatletter
\providecommand \@ifxundefined [1]{%
 \@ifx{#1\undefined}
}%
\providecommand \@ifnum [1]{%
 \ifnum #1\expandafter \@firstoftwo
 \else \expandafter \@secondoftwo
 \fi
}%
\providecommand \@ifx [1]{%
 \ifx #1\expandafter \@firstoftwo
 \else \expandafter \@secondoftwo
 \fi
}%
\providecommand \natexlab [1]{#1}%
\providecommand \enquote  [1]{``#1''}%
\providecommand \bibnamefont  [1]{#1}%
\providecommand \bibfnamefont [1]{#1}%
\providecommand \citenamefont [1]{#1}%
\providecommand \href@noop [0]{\@secondoftwo}%
\providecommand \href [0]{\begingroup \@sanitize@url \@href}%
\providecommand \@href[1]{\@@startlink{#1}\@@href}%
\providecommand \@@href[1]{\endgroup#1\@@endlink}%
\providecommand \@sanitize@url [0]{\catcode `\\12\catcode `\$12\catcode `\&12\catcode `\#12\catcode `\^12\catcode `\_12\catcode `\%12\relax}%
\providecommand \@@startlink[1]{}%
\providecommand \@@endlink[0]{}%
\providecommand \url  [0]{\begingroup\@sanitize@url \@url }%
\providecommand \@url [1]{\endgroup\@href {#1}{\urlprefix }}%
\providecommand \urlprefix  [0]{URL }%
\providecommand \Eprint [0]{\href }%
\providecommand \doibase [0]{https://doi.org/}%
\providecommand \selectlanguage [0]{\@gobble}%
\providecommand \bibinfo  [0]{\@secondoftwo}%
\providecommand \bibfield  [0]{\@secondoftwo}%
\providecommand \translation [1]{[#1]}%
\providecommand \BibitemOpen [0]{}%
\providecommand \bibitemStop [0]{}%
\providecommand \bibitemNoStop [0]{.\EOS\space}%
\providecommand \EOS [0]{\spacefactor3000\relax}%
\providecommand \BibitemShut  [1]{\csname bibitem#1\endcsname}%
\let\auto@bib@innerbib\@empty
\bibitem [{cor(2021)}]{corning2021}%
  \BibitemOpen
  \href {https://www.corning.com/media/worldwide/cdt/documents/EAGLE%20XG_PI%20Sheet_2021.pdf} {\bibinfo {title} {{Corning Eagle XG glass datasheet}}} (\bibinfo {year} {2021})\BibitemShut {NoStop}%
\bibitem [{\citenamefont {Powell}\ \emph {et~al.}(1966)\citenamefont {Powell}, \citenamefont {Ho},\ and\ \citenamefont {Liley}}]{powell1966thermal}%
  \BibitemOpen
  \bibfield  {author} {\bibinfo {author} {\bibfnamefont {R.}~\bibnamefont {Powell}}, \bibinfo {author} {\bibfnamefont {C.~Y.}\ \bibnamefont {Ho}},\ and\ \bibinfo {author} {\bibfnamefont {P.~E.}\ \bibnamefont {Liley}},\ }\href@noop {} {\emph {\bibinfo {title} {Thermal conductivity of selected materials}}},\ Vol.~\bibinfo {volume} {8}\ (\bibinfo  {publisher} {US Department of Commerce, National Bureau of Standards Washington, DC},\ \bibinfo {year} {1966})\BibitemShut {NoStop}%
\bibitem [{\citenamefont {Matula}(1979)}]{matula1979electrical}%
  \BibitemOpen
  \bibfield  {author} {\bibinfo {author} {\bibfnamefont {R.~A.}\ \bibnamefont {Matula}},\ }\bibfield  {title} {\bibinfo {title} {Electrical resistivity of copper, gold, palladium, and silver},\ }\href@noop {} {\bibfield  {journal} {\bibinfo  {journal} {Journal of Physical and Chemical Reference Data}\ }\textbf {\bibinfo {volume} {8}},\ \bibinfo {pages} {1147} (\bibinfo {year} {1979})}\BibitemShut {NoStop}%
\bibitem [{\citenamefont {Ceccarelli}\ \emph {et~al.}(2020)\citenamefont {Ceccarelli}, \citenamefont {Atzeni}, \citenamefont {Pentangelo}, \citenamefont {Pellegatta}, \citenamefont {Crespi},\ and\ \citenamefont {Osellame}}]{ceccarelli2020low}%
  \BibitemOpen
  \bibfield  {author} {\bibinfo {author} {\bibfnamefont {F.}~\bibnamefont {Ceccarelli}}, \bibinfo {author} {\bibfnamefont {S.}~\bibnamefont {Atzeni}}, \bibinfo {author} {\bibfnamefont {C.}~\bibnamefont {Pentangelo}}, \bibinfo {author} {\bibfnamefont {F.}~\bibnamefont {Pellegatta}}, \bibinfo {author} {\bibfnamefont {A.}~\bibnamefont {Crespi}},\ and\ \bibinfo {author} {\bibfnamefont {R.}~\bibnamefont {Osellame}},\ }\bibfield  {title} {\bibinfo {title} {Low power reconfigurability and reduced crosstalk in integrated photonic circuits fabricated by femtosecond laser micromachining},\ }\href@noop {} {\bibfield  {journal} {\bibinfo  {journal} {Laser \& Photonics Reviews}\ }\textbf {\bibinfo {volume} {14}},\ \bibinfo {pages} {2000024} (\bibinfo {year} {2020})}\BibitemShut {NoStop}%
\end{thebibliography}%

\end{document}


\title{Integrated thermo-optic phase shifters for laser-written photonic circuits operating at cryogenic temperatures}

\author{Francesco Ceccarelli}
\affiliation{\ifn}

\author{Jelena V. Rakonjac}
\affiliation{\icfo}

\author{Samuele Grandi}
\affiliation{\icfo}

\author{Hugues de Riedmatten}
\affiliation{\icfo}
\affiliation{\icrea}

\author{Roberto Osellame}
\affiliation{\ifn}

\author{Giacomo Corrielli}
\thanks{Corresponding author: giacomo.corrielli@cnr.it}
\affiliation{\ifn}

 
\section*{Supplementary material: Analytical modeling and scaling rule}
\subsection{TOPS with no pillars}
Here we propose a simple analytical model to calculate the temperature variation along the bridge induced by a power dissipation $P$ in vacuum and we thoroughly study its applicability and possible limitations. Accordingly to the main text, let $b$, $h_1$ and $L_u$ be the width, height and length of the bridge, respectively, and $t$ is the lateral width of the resistors. Moreover, let us assume uniform, isotropic and linear media (i.e. metal heater and glass bridge), which basically means assuming a constant scalar value for the thermal conductivities and uniform power dissipation along the bridge. Under these assumptions, the temperature distribution can be calculated by solving the steady-state heat equation, namely:
\begin{equation}\label{eqheat}
    k \nabla^2 T = -\sigma.
\end{equation}
where $k$ is the thermal conductivity of the medium and $\sigma$ is the power dissipation density measured in W/m$^3$. Generally speaking, such an equation should be solved separately in each medium by considering the dependence of $T$ on all the coordinates $x$, $y$ and $z$. However, if we neglect the presence of the pillars and consider perfect insulation around the bridge, heat can diffuse only in the direction parallel to the waveguide and, thus, the temperature gradient will have only the $x$ component. This also means that glass and metal can be considered as a whole medium featuring an effective thermal conductivity defined as:
\begin{equation}\label{eqheat1D}
    k_{eff} = \frac{k_{glass} b h_1 + k_{gold} t h_{3}}{b h_1 + t h_{3}} \simeq \frac{k_{glass} b h_1 + k_{gold} t h_{3}}{b h_1},
\end{equation}
where $k_{glass}$ = 1.09 W/mK \cite{corning2021} is the thermal conductivity of our glass, $k_{gold}$ = 315 W/mK \cite{powell1966thermal} is the thermal conductivity of gold and $h_{3}$ = 100 nm is the thickness of the metal film. Thanks to the larger cross section of the glass bridge with respect to the metal counterpart, it is possible to simplify this equation by assuming $k_{eff} \simeq k_{glass}$, i.e. all the heat is flowing through the glass. Such a simplification leads to an error of less than 10\% on the effective thermal conductivity. However, this estimation is only an upper bound that might represent a strong overestimation of the error. Indeed, if we consider the fact that the actual electrical conductivity of the thin gold layer is about 5.2 MS/m, almost an order of magnitude lower with respect to the bulk reference (i.e. 45.2 MS/m \cite{matula1979electrical}), assuming that the thermal conductivity of the thin gold layer is following a similar trend looks reasonable. This leads to a strong simplification of the heat equation, that now reads:
\begin{equation}\label{eqheat1Dapprox}
    k_{glass} \frac{d^2T}{dx^2} = -\frac{P}{b h_1 L_u}.
\end{equation}
The general integral satisfying this relation is given by:
\begin{equation}\label{eqgeneral}
    \Delta T(x) = T(x) - T_0 = -\frac{P}{2 k_{glass} b h_1 L_u}x^2 + Bx + C,
\end{equation}
where $\Delta T$ is the temperature variation with respect to the global temperature $T_0$ of the substrate, while $B$ and $C$ are arbitrary constants that must be determined by imposing suitable conditions at the boundary of the domain, i.e. at the beginning and at the end of the bridge. As soon as the heat is collected by the bulk regions at the ends of the bridge, the heat flux will promptly spread and, therefore, the temperature gradient will rapidly fade out. Such considerations leads us to assume a negligible increase of the temperature and, therefore, Dirichlet boundary conditions:  
\begin{equation}\label{eqboundary}
    \begin{cases}
        \Delta T\left(-\frac{L_u}{2}\right) = 0,\\
        \Delta T\left(\frac{L_u}{2}\right) = 0.
    \end{cases}
\end{equation}
As a result, the temperature distribution $\Delta T(x)$ becomes:
\begin{equation}\label{eqtemp}
    \Delta T(x) = \frac{P}{2k_{glass} b h_1 L_u}\left(\frac{L_u^2}{4}-x^2\right).
\end{equation}

\begin{figure}[t]
    \centering
    \includegraphics[width=1\linewidth]{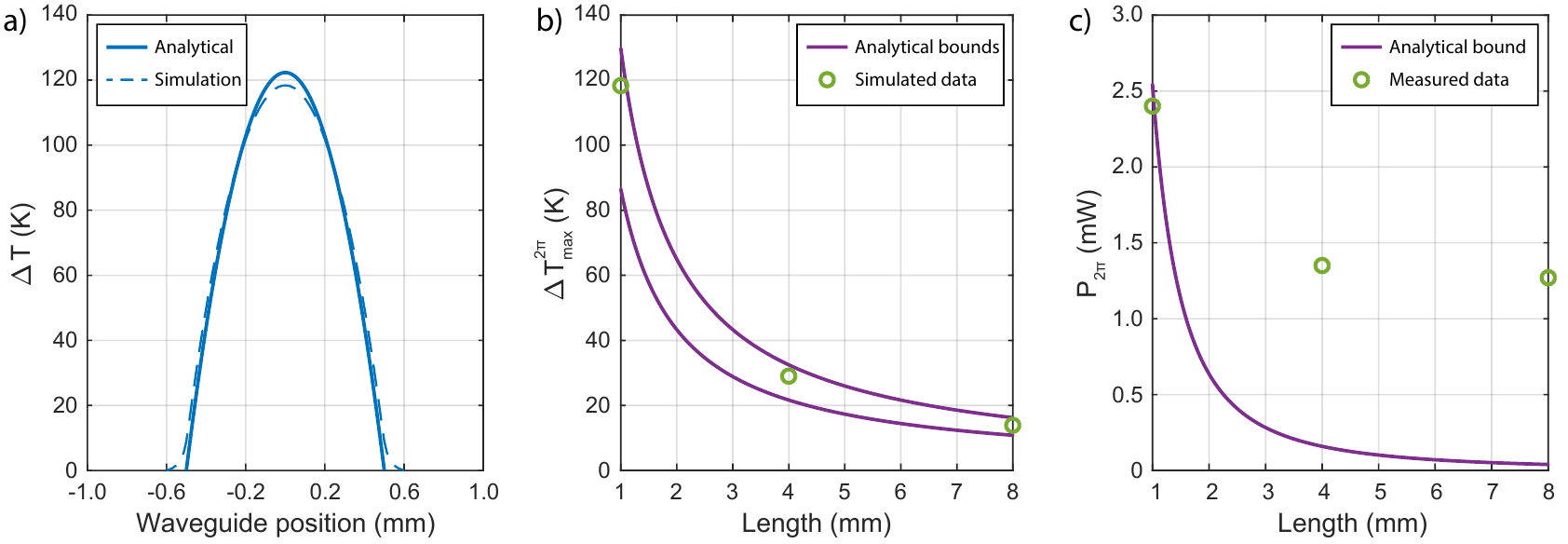}
    \caption{(a) Temperature variation $\Delta T$ as a function of the position along the bridge for MZI 1: comparison between the analytical model and the simulation. (b) Maximum temperature reached along the bridge for a $2\pi$ phase shift as a function of the TOPS length. Simulated data reported in the main text are compared to the analytical bounds derived here. (c) Power dissipation needed for a $2\pi$ phase shift as a function of the TOPS length. The measured data at room temperature reported in the main text are compared to the lower analytical bound derived here.} 
    \label{FigA1}
\end{figure}

The analytical temperature profile calculated for MZI 1 ($L_u$ = 1 mm) in the case of $P=P^A_{2\pi}$ is reported in figure \ref{FigA1}(a) along with the simulated one already reported in figure 4(b) of the main text. Despite the simplicity of our analytical model, the maximum difference between these two curves is always less than a few degrees, thus demonstrating the goodness of our hypotheses. The main origin of the small differences between the two curves can be ascribed to the choice of considering a thermal conductivity that does not depend on the temperature and a perfect thermal contact at the ends of the bridge. However, such a good approximation can not be reached also for MZI 2 and 3. Indeed, the simulated profile of figure 4(b) of the main text is qualitatively very far from being parabolic. In addition, this is evident also from a more quantitative perspective when we calculate the maximum temperature variation
\begin{equation}\label{eqtemp_mean}
    \Delta T_{max} = T(0) = \frac{P L_u}{8k_{glass} b h_1},
\end{equation}
and the mean temperature variation 
\begin{equation}\label{eqtemp_max}
    \Delta T_M = \frac{1}{L_u} \int_{-\frac{L_u}{2}}^{\frac{L_u}{2}} \Delta T(x) dx = \frac{P L_u}{12k_{glass} b h_1} = \frac{2}{3} \Delta T_{max},
\end{equation}
which are both directly proportional to the length $L_u$, a dependence not compatible with the simulation data reported in figure 4 of the main text. We ascribe such a large discrepancy to the presence of the glass pillars, which introduce dissipation channels that can not be ignored.

The model developed so far links temperature and dissipated power in compact devices that do not require pillars, but it does not take into account information about the actual phase difference induced between the arms of the interferometer. This quantity can be calculated as reported in \cite{ceccarelli2020low}, namely:
\begin{equation}\label{eqphase}
    \phi = \frac{2\pi}{\lambda} c_{to} \Delta T_M L_u.
\end{equation}
From this relation it is possible to set $\phi = 2\pi$ and derive physical parameters that are very important for the designer such as the $2\pi$ maximum temperature
\begin{equation}\label{eqt2pi}
    \Delta T^{2\pi}_{max} = \frac{3}{2}\Delta T^{2\pi}_{M} = \frac{3\lambda}{2c_{to}L_u},
\end{equation}
which must be always considered to guarantee a stable and reliable phase shift over time, and the $2\pi$ power dissipation 
\begin{equation}\label{eqp2pi}
    P_{2\pi} = \frac{12 \lambda k_{glass} b h_1}{c_{to} L_u^2},
\end{equation}
which sets the maximum power needed by a TOPS in a MZI-based mesh. Two main differences can be appreciated with respect to MZIs operating in standard pressure conditions \cite{ceccarelli2020low}: (i) the TOPS witnesses a maximum temperature that is a factor 3/2 higher due to the non-uniform temperature distribution $\Delta T(x)$ along the bridge, although it shows the same inversely proportional dependence on the length $L_u$ that is typical in standard conditions. (ii) The TOPS dissipates a power that is inversely proportional to the squared length $L_u^2$ and, thus, power dissipation and miniaturization can not be chosen independently as happens in standard conditions.

\subsection{TOPS with pillars}
These results are not only useful to understand the scaling rules in compact devices that do not require pillars, but also to obtain bounds for the physical quantities involved in the design process of longer TOPSs featuring pillars. First, the $2\pi$ maximum temperature predicted by equation \ref{eqt2pi} represents an upper bound for the temperature, while the $2\pi$ mean temperature $\Delta T_M^{2\pi} = \lambda/c_{to}L_u$ represents the lower bound that is reached only for the limit of no insulation underneath the waveguide (i.e. uniform temperature along the bridge, as in standard pressure conditions). Mathematically speaking:
\begin{equation}\label{eqt_bound}
     \frac{\lambda}{c_{to}L_u} \leq \Delta T^{2\pi}_{max} \leq \frac{3\lambda}{2c_{to}L_u}.
\end{equation}
Figure \ref{FigA1}(b) includes both the two bounds and the $2\pi$ maximum temperatures simulated and reported in the main text for all the three MZIs. Interestingly, the temperature span within the two bounds is quite narrow, thus meaning that both full and zero insulation hypothesis can be used to get reasonable results. On the other hand, equation \ref{eqp2pi} represents a lower bound for the $2\pi$ power dissipation, namely:
\begin{equation}\label{eqp_bound}
    P_{2\pi} \geq \frac{12 \lambda k_{glass} b h_1}{c_{to} L_u^2}.
\end{equation}
Figure \ref{FigA1}(c) includes both the bound and the $2\pi$ powers measured at room temperature and reported in the main text for all the three MZIs. Interestingly, these values are far from the bound set by equation \ref{eqp_bound}, thus leaving large room for improvement provided that reducing either the section or the number of the pillars is compatible with the mechanical stability of the structures. Both figures \ref{FigA1}(b) and (c) consider $c_{to}$ = 7E-6 K$^{-1}$ and $\lambda$ = 606 nm. 

\bibliography{bibliography.bib}